\documentclass[]{article}
\usepackage{graphicx}
\usepackage{color}
\usepackage{listings}
\usepackage{fullpage}
\usepackage{amsmath}
\usepackage[utf8x]{inputenc}
\usepackage{import}
\usepackage{setspace}
\usepackage{authblk}
\usepackage{hyperref}
\usepackage{float}
\definecolor{lightgray}{gray}{0.5}
\usepackage{setspace}
\usepackage{color}
\usepackage[super]{nth}
\usepackage[font={small,sf}, singlelinecheck=false]{caption}

\usepackage{epigraph}
\usepackage{changepage}

\title{Neural dynamics of cognitive control: Current tensions and future promise}

\author[1,2,3]{Dale Zhou}
\author[4,5,6]{Danielle Cosme}
\author[4,7]{Yoona Kang}
\author[4,5,6]{Ovidia Stanoi}
\author[4,8]{David M. Lydon-Staley}
\author[9]{Peter J. Mucha}
\author[4,5,6]{Emily B. Falk}
\author[10]{Kevin N. Ochsner}
\author[11-17]{Dani S. Bassett}

\affil[1]{Department of Neurobiology and Behavior, University of California, Irvine}
\affil[2]{University of California, Irvine, Center for the Neurobiology of Learning and Memory, Qureshey Research Laboratory, Irvine, 92697, United States}
\affil[3]{Department of Cognitive Sciences, University of California, Irvine}
\affil[4]{Annenberg School for Communication, University of Pennsylvania}
\affil[5]{Department of Psychology, University of Pennsylvania}
\affil[6]{Marketing Department, Wharton School, University of Pennsylvania}
\affil[7]{Rutgers, The State University of New Jersey}
\affil[8]{Leonard Davis Institute of Health Economics, University of Pennsylvania}
\affil[9]{Department of Mathematics, Dartmouth College}
\affil[10]{Department of Psychology, Columbia University}
\affil[11]{Department of Bioengineering, School of Engineering and Applied Sciences, University of Pennsylvania}
\affil[12]{Department of Physics \& Astronomy, College of Arts and Sciences, University of Pennsylvania}
\affil[13]{Department of Neurology, Perelman School of Medicine, University of Pennsylvania}
\affil[14]{Department of Electrical \& Systems Engineering, School of Engineering and Applied Sciences, University of Pennsylvania}
\affil[15]{Department of Psychiatry, Perelman School of Medicine, University of Pennsylvania}
\affil[16]{Santa Fe Institute}
\affil[17]{To whom correspondence should be addressed: dsb@seas.upenn.edu}

\begin{document}

\date{}
\maketitle

\newpage

\begin{abstract}
    Cognitive control is a suite of processes that helps individuals pursue goals despite resistance or uncertainty about what to do. Deficits of cognitive control underlie compulsive or risky behavior, as well as other clinical challenges associated with difficulties in regulating impulses, attention, thoughts, and feelings. Although cognitive control has been extensively studied as a dynamic feedback loop of perception, valuation, and action, it remains incompletely understood as a cohesive dynamic and distributed neural process. Here, we critically examine the history of and advances in the study of cognitive control, including how metaphors and cultural norms of power, morality, and rationality are intertwined with definitions of control, to consider holistically how different models explain which brain regions act as controllers. Controllers, the source of top-down signals, are typically localized in regions whose neural activations implement elementary component processes of control, including conflict monitoring and behavioral inhibition. Top-down signals from these regions guide the activation of other task-specific regions, biasing them towards task-specific activity patterns. A relatively new approach, network control theory, has roots in dynamical systems theory and systems engineering. This approach can mathematically show that controllers are regions with strongly nested and recurrent anatomical connectivity that efficiently propagate top-down signals, and precisely estimate the amount, location, and timing of signaling required to bias global activity to task-specific patterns. Importantly, the theory converges with established findings, provides new mathematical tools and intuitions for understanding control loops across levels of analysis, and naturally produces graded predictions of control across brain regions and modules of psychological function that have been unconsidered, marginalized, or indirectly linked. We describe how psychological and network control approaches converge and diverge, noting directions for future integration that could strengthen and sharpen our understanding and predictions of how the brain instantiates cognitive control.
\end{abstract}

\section{Introduction}

Meeting the demands of everyday life requires the ability to flexibly pursue all manner of goals, overcoming obstacles and resisting distractions along the way. In psychology, this ability is known as cognitive control, which is as crucial for single-minded striving toward one end as it is for flexibly changing aims~\cite{carver2001self}. Psychological definitions of control frame it as a suite of processes for prioritizing the pursuit of goals that may conflict with competing thoughts, emotions, impulses, motivations, desires, other goals, and behaviors. For example, cognitive control underlies the ability to inhibit responses that conflict with goals---representations of long-term and momentary outcomes---while also pursuing responses that align with them. To manage such conflicts, several subprocesses are thought to be involved, including conflict monitoring, cognitive flexibility, task switching, working memory, selective attention, inhibitory control, and learning \cite{monsell2003task, egner2005cognitive, munakata2011unified, miyake2012nature, jamadar2015task, braunstein2017explicit, friedman2017unity, eisenberg2019uncovering}. Some goals, reached via habitual or automatic behaviors (e.g., thoroughly learned tasks), do not need cognitive control and using it to guide these behaviors would be counterproductive or an inefficient use of effortful control resources~\cite{poldrack2005neural, diamond2013executive, neal2013people, amer2016cognitive, bustamante2021learning}. Pursuing some goals rather than others means weighing the value of allocating resources and time. Given that time is finite, as are the resources one can devote to control, this points towards the necessity of computing the value of exerting control~\cite{yee2018interactions, shenhav2016dorsal, stanovich2020humans, touroutoglou2020tenacious}. Once seen as a fatigable willpower akin to tired muscles~\cite{ribot1911voluntary}, control energy is now understood in terms of allocating mental resources including attention, working memory, motivation, and effort~\cite{kool2018mental}. 

The importance of cognitive control is evident in its broad conceptual scope and applicability to a wide range of behavioral contexts, from the kitchen and cashier to the classroom, clinic, and courtroom, and even to one's own conscience. Conceptually, cognitive control is intertwined with broader behavioral constructs such as executive function, self-regulation, self-discipline, willpower, volition, delayed gratification, risk-taking, and emotion regulation \textbf{(Figure \ref{fig1})}~\cite{duckworth2011meta, mischel2011willpower, fujita2011conceptualizing, diamond2013executive, burman2015meanings, nigg2017annual, inzlicht2021integrating}. Alterations in control function are hallmarks of many neurological and psychiatric conditions, contributing to distress and challenges to daily function related to loss of agency and autonomy~\cite{arnsten2012neurobiological, mcteague2017identification, barch2018systems, romer2021regulatory}. Aspects of cognitive control have been analogized to a ``moral muscle'' related to a strength of will to resist temptations~\cite{park2004strengths, duckworth2007grit, hofmann2018morality, wente2020developmental, speer2022cognitive}, and research on the developmental trajectory of cognitive control has influenced legal standards for personal responsibility \cite{muller2015development, luna2012relevance, luna2016adolescent}. Finally, control is thought to have a causal power over actions that paves the way for desired outcomes in school, health, finances, and criminal justice with consequences for individuals and societies~\cite{duckworth2005self, blair2007relating, heatherton2011cognitive, moffitt2011gradient, mischel2011willpower}. These applications highlight a powerful normative worldview about cognitive control historically rooted in beliefs about power, morality, and rationality~\cite{ribot1911voluntary, gennara2023more, hofmann2018morality}, which require critical reexamination~\cite{lydon2014adolescent, bermudez2024believe}.

\begin{figure}
\begin{center}
 \includegraphics[width=\columnwidth]{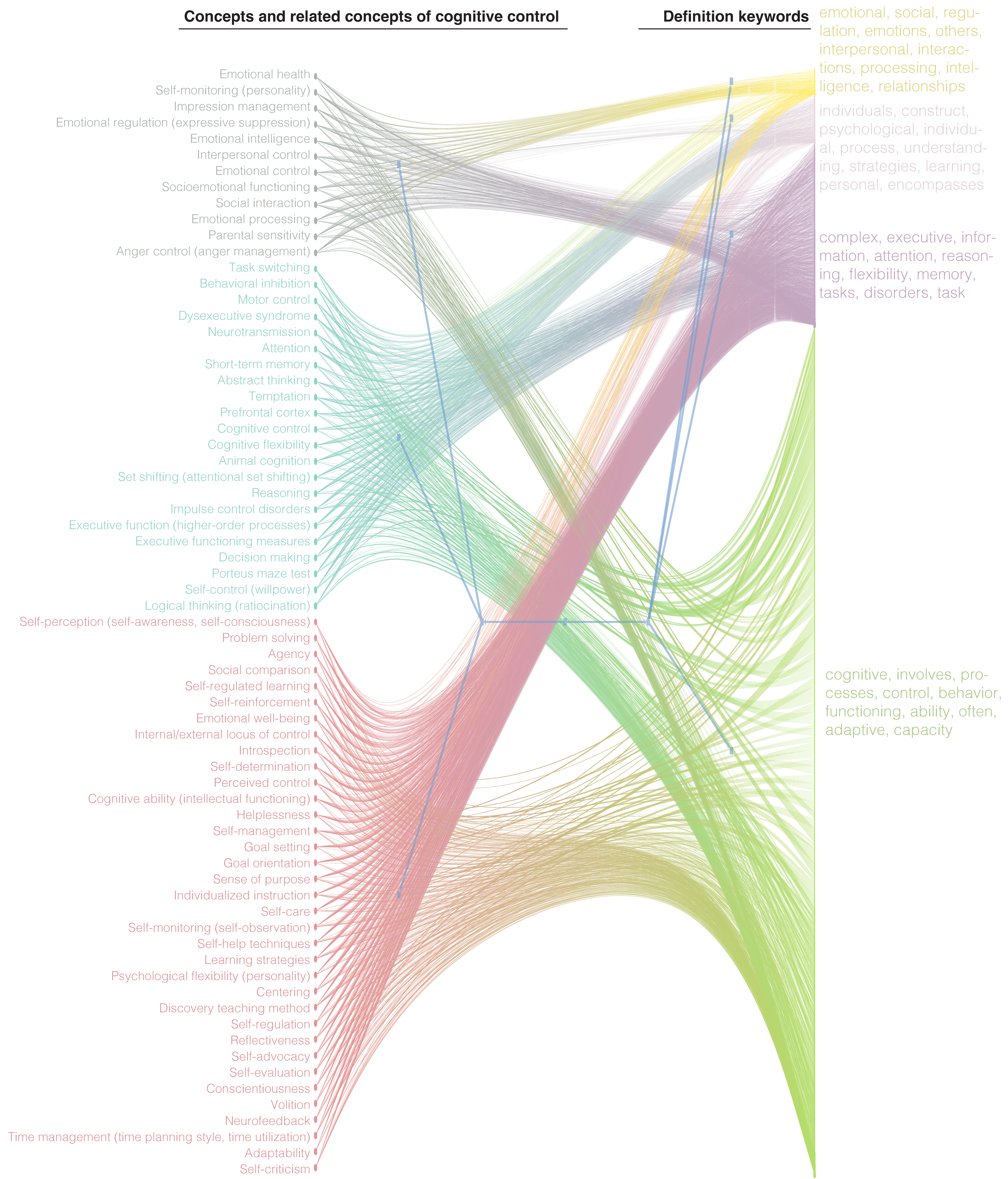}
\end{center}
\caption{\textbf{Self-regulation concepts.} Conflicting and fragmented working definitions of cognitive control remain a major challenge for developing cohesive theories, with many concepts used interchangeably despite continuous calls for clarity. Inconsistency is unavoidable as a force of scientific progress. In addition to pursuing more ecologically relevant tasks, clearer definitions, and consensus, understanding may come from embracing the plurality of basic scientific and applied goals across disciplinary boundaries. The American Psychological Association Thesaurus of Psychological Index Terms®}
\label{fig1}
\end{figure}

\begin{figure}[t]\ContinuedFloat
\caption{provides concepts directly related to the concepts of ``cognitive control'' and ``self-regulation,'' as well as the terms' definitions~\cite{APAThesaurusDigital, burman2015meanings}. Colors indicate groupings of the concept nodes based on community detection of the concept-by-keyword co-occurrence network~\cite{gerlach2018network}. Concept nodes are directly or indirectly related to the ``cognitive control'' and ``self-regulation'' concepts, such as self-control, executive function, and agency. For the keyword nodes, we show the most frequent words within each group. Edges indicate co-occurrence. The blue tree network indicates hierarchical community structure. See Appendix \ref{fig1_method} for methodological details.}
\end{figure}

Although much progress has been made towards characterizing cognitive control at multiple levels of analysis, many open questions remain. At present, working descriptions of cognitive control make different assumptions about what behaviors to focus on, the kinds of processes or representations involved, and how neural substrates instantiate control processes and representations.  As such, questions remain about the goals and behaviors that exemplify ``good'' cognitive control, whether control processes and abilities are domain specific vs. domain general, and whether and how neural systems implement putative processes and mental representations. Varied and disparate working definitions of cognitive control represent major efforts from diverse perspectives to address these and related questions \textbf{(Figure \ref{fig2})} ~\cite{burman2015meanings, kotabe2015integrating, nigg2017annual, milyavskaya2019many, eisenberg2019uncovering, bunge2024should}.

Although psychologists mostly agree on what kinds of behaviors require cognitive control, conceptualizations of underlying processes still largely involve mapping individual component processes onto individual neural systems. This approach has identified various kinds of putative systems underlying control functions, such as working memory or conflict monitoring. However, this componential approach is difficult to reconcile with recent advances in computational, systems, and network neuroscience that emphasize parallel processing with dynamic and distributed representations. As such, current models of control can seem more akin to a chimeric list of behaviors and component parts rather than a coherent and holistic explanation of how complex behaviors emerge from control processes implemented by interacting neural systems.

By contrast, a neurobiological process account that highlights cognitive control as a distributed and dynamic network function could strengthen cohesion across levels of analysis~\cite{eisenreich2017control, pessoa2023entangled, noble2023tip}. A distributed account explicitly models the collective contribution of regions, including those not yet recognized by current paradigms. A dynamic account centers how control operates from moment-to-moment as a function of the evolution of collective activity, rather than only as a function of a terminal state where regions (co-)activate past a threshold. A network account models the brain as a complex system of units (nodes) interconnected by physical pathways (edges), and provides tools to formally quantify collective dynamics across too many regions to grasp without mathematical abstraction. At the intersection of distributed, dynamic, and network models is network control theory. 

Similar to other major advances~\cite{shenhav2013expected}, the value of network control theory does not lie primarily in the novelty of its core components---nodes, edges, dynamics, and control targets---but in its explicit formalization of these elements in an interpretable mathematical framework. This endeavor makes it possible to precisely characterize the kinds of control a brain region can exert over its activity toward both known and yet unknown control targets. Network control theory helps synthesize many key ideas in the neuroscience of cognitive control in a precise and mathematically coherent account, linking the processes underlying regulation, development, hierarchy, and cost~\cite{cohen1990control, desimone1995neural, miller2000prefontral, badre2008cognitive, cole2017control, badre2018frontal, luna2015integrative, marek2015contribution, crone2017neural, shenhav2013expected, cools2015cost, westbrook2020dopamine}, while producing new predictions for unconsidered control components, processes, and targets. For example, network control theory advances understanding of cognitive control by identifying key brain regions and their development, predicting when and how much control is required, and extending insights across species. It improves behavioral and psychiatric prediction beyond simpler network metrics (e.g. the strength of a region's connectivity), and offers a framework for studying and modeling neurological and psychiatric (pharmacological, neurostimulation, psychological) interventions.

\subsection*{Goals of the review}
Philosophers and scientists have theorized extensively about how people form goals, make decisions to support them, and flourish through attaining them. What constitutes a good life? Where do goals come from? How do we know if we are achieving them? Defining cognitive control becomes more difficult when both scientific and cultural frameworks constrain how we think about it, historically conflating control with willpower, relying on metaphors about social structures of power, and focusing on individual agency rather than collectives. When theories of cognitive control are shaped by the very social and psychological forces they aim to explain, the risk is that beliefs, values, and assumptions that have become second nature are embedded in our models of nature. How might we move beyond some of these assumptions and expand our ability to see how life actually works and as importantly how it could work? Taking a step back to describe how energy flows in the brain and moves people through decisions and actions with the mathematical language of network control theory offers new insight into how control works and how it might be leveraged to promote flourishing ~\cite{medaglia2019clarifying, lynn2019physics, badre2020brain}.

We first define key ideas across psychological and neuroscientific theories of cognitive control, and then highlight tensions which motivate a new conceptualization of cognitive control. For more detail on core psychological concepts, we point readers to literature on the psychology of goal pursuit and self-regulation~\cite{gollwitzer2013history, inzlicht2021integrating,  brandstatter2022persistence}, computational models of cognitive control \cite{yee20, yee2018computational}, syntheses of cognitive control across regions and units of analysis \cite{satpute2012neuroscience, egner2017wiley, gratton2018dynamics, badre2022task}, the role of specialized brain regions (e.g., prefrontal cortex \cite{miller2000prefontral, friedman2022role, menon2022role}, dorsal anterior cingulate cortex \cite{shenhav2013expected, clairis2023debates}, and insular regions~\cite{menon2010saliency, uddin2021cognitive}) in allocating control~\cite{silvestrini2023integrative}, component processes across cognitive control and executive function tasks \cite{miyake2012nature, niendam2012meta, diamond2013executive, friedman2017unity, jamadar2015task}, and the neurodevelopment of cognitive control \cite{somerville2010developmental, luna2010has, munakata2012developing, luna2015integrative, crone2017neural}.

The next section conveys mathematical intuitions underlying network control theory, showing what is unique about the formulation with no assumption of prerequisite mathematical background. For more extensive mathematical treatments of the topic, see literatures on the general theory and applications \cite{motter2015networkcontrology, liu2016control, d2023controlling}, applications to neuroscience \cite{tang2018colloquium,lynn2019physics, srivastava2020models}, didactic textbook chapters \cite{kim2020linear, brunton2022data}, practical tutorials and code \cite{karrer2020practical, parkes2023using}, application to animal models including directed connectomes of \emph{C. elegans} and mice \cite{towlson2018caenorhabditis, parkes2023using}, and applications to perturbing and predicting neural dynamics \cite{medaglia2017mind, shenoy2021measurement, srivastava2022expanding}. In the final section, we outline how each approach contributes to the other and potential future efforts to interdigitate the two constructs. When we refer to the psychological perspective of control, we use the term ``cognitive control,'' whereas we refer to the network science perspective, we use the term ``network control.'' When referring to the broader concept of self-regulation, we simply say ``control.''

\section{Cognitive Control: Constructs and Correspondence}

\subsection*{The psychological construct of cognitive control}

\begin{adjustwidth}{2em}{0pt}
\textit{``If in general we class all springs of action as propensities on the one hand and ideals on the other, the sensualist never says of his behavior that it results from a victory over his ideals, but the moralist always speaks of his as a victory over his propensities. [...] And if a brief definition of ideal or moral action were required, none could be given which would better fit the appearances than this: It is action in the line of the greatest resistance.''}

\hfill ------ Ref. \cite{james1890principles}
\end{adjustwidth}

\subsubsection*{Historical origins}

Why would anyone act contrary to their own goals and ideals~\cite{davidson2001weakness}? It has long been recognized that achieving feats of self-control is difficult because different mental, emotional, and motivational processes conflict with one another. Plato conceptualized the mind as a team of horses representing passion and appetite controlled by a charioteer representing reason~\cite{plato1952phaedrus}. Without the charioteer, the horses run wild and the mind cannot pursue virtuous purposes. Continuing in this contest between ``dual processes,'' the study of cognitive control has historically been framed as a conflict between uncontrolled and controlled aspects of the mind: automatic vs. deliberate; hot vs. cold; reactionary vs. willful; stimulus-driven vs. goal-driven; impulsive vs. inhibited; id vs. ego; distracted vs. attentive; off-task vs. on-task, and habitual vs. effortful. Such contests are frequently moralized as conflict between inner demons versus better angels, wherein the victory of virtues over vices drives human progress~\cite{elias1939civilizing, read1999mixing, pinker2012better, heckhausen2018historical, hofmann2018morality}. 

Over a century ago, at the dawn of psychology, control was thought to tip the scales in these conflicts. This idea was captured by a simple yet enduring formula: cognitive control energy or effort $E$ decides the balance between habitual instincts, impulses, motives, pleasures, and pains as propensities $P$ and our ideals $I$~\cite{james1890principles}. $E$ is needed when $I < P$; that is, when habits override ideal goals. Cognitive control allows for $I + E > P$. 

\subsubsection*{Recent advances}

Although oversimplified, this basic formulation has been expanded or rearranged to explain many aspects of control behavior. $E$ is not a unitary resource of will but a decomposable process with multiple elementary components~\cite{nee2007interference, nee2008dissociable}. $E$ is partly about weighing the value of alternatives and monitoring progress~\cite{miller1960plans, daw2005uncertainty, mcguire2013rational, yee20, berkman2017self, yee2018computational, lieder2018rational, musslick2021rationalizing}, a motivated and context-sensitive valuation of how to allocate time and let go of lost opportunities. This does not merely rebrand willpower as willingness to apply effort; this $E$ is a subjective valuation that weighs the costs and benefits of $I$ as contexts change, tracking when $I$ promises greater reward, $I$ is more probable, and reduced $E$ can lead to better control outcomes~\cite{padmala2011reward, kool2017cognitive, bustamante2021learning}. $E$ can also act as a divisor to $P$, rearranging the inequality to bring $E$ to the other side as an inhibitory gate so that $I > \frac{P}{E}$~\cite{heatherton2011cognitive, munakata2011unified, miyake2012nature, shenhav2013expected, kotabe2015integrating, bunge2024should}. Other components amplify or diminish $E$, such as emotions and motivations, and $P$ is a function of many states and traits, such as risk taking, sensation seeking, and conscientiousness~\cite{shackman2011integration, zelazo2012hot, ochsner2012functional, shenhav2013expected, pessoa2017cognitive, eisenberg2019uncovering, enisman2024relative}. Taken together, $E$ is not a unitary input to tip the balance between (or among) habits and goals, but is a series of calculated choices shaped by separate but interacting component processes~\cite{wood2022habits}.

\subsubsection*{Component processes}

\begin{figure}
\begin{center}
 \includegraphics[width=\columnwidth]{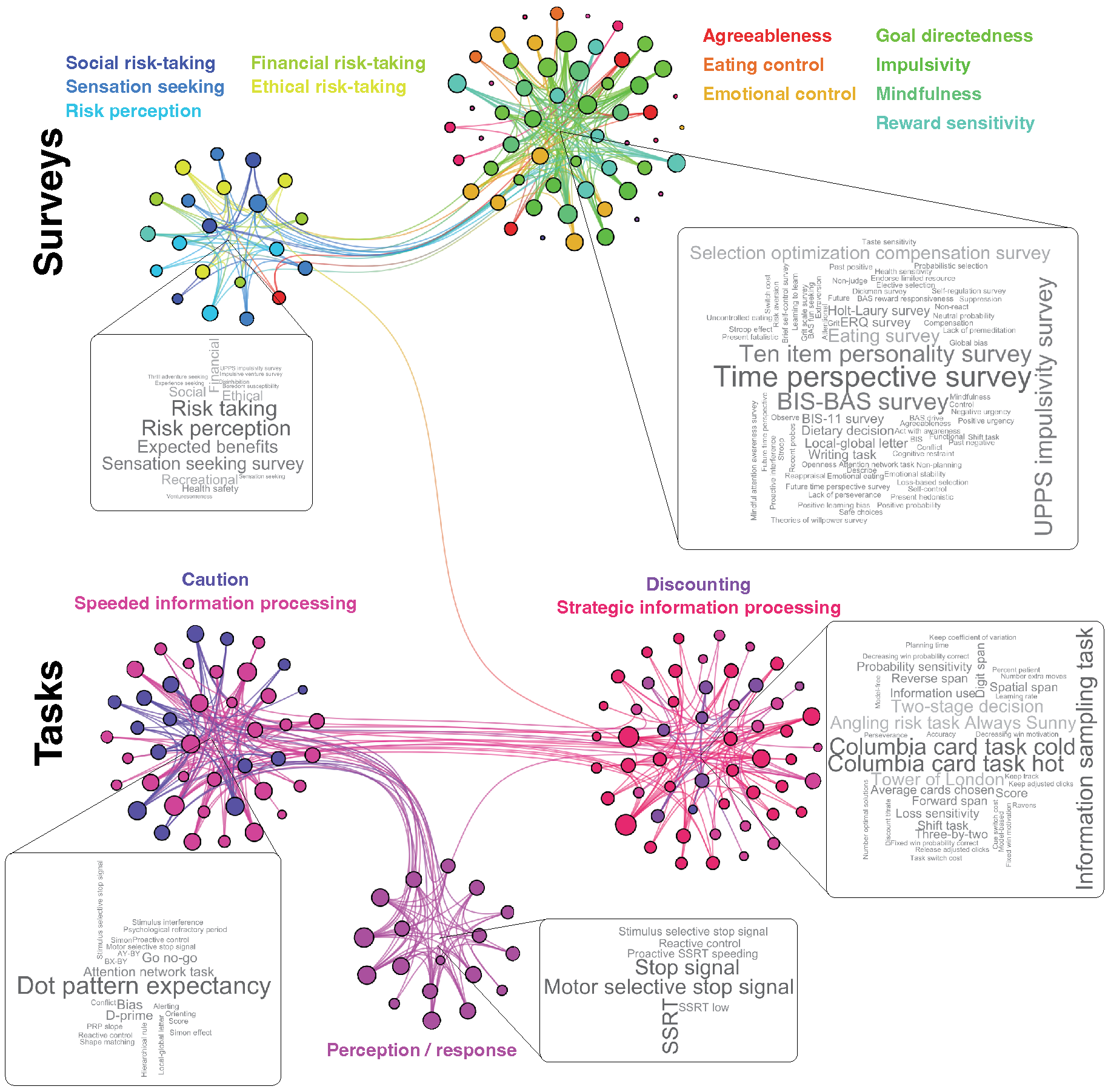}
\end{center}
\caption{\textbf{Self-regulation measurements from tasks and surveys.} Self-regulation often refers to processes that support goal pursuit broadly, and may benefit from but do not necessarily require certain forms of cognitive control~\cite{fujita2011conceptualizing}. Current survey and task measures, including cognitive control constructs, exhibit limited coherence (weak connections between task and survey modalities and within conceptual domains). A ``cognitive ontology'' is visualized using 129 task and 64 survey dependent variables reproduced from previously published data~\cite{eisenberg2019uncovering}. Nodes are defined as dependent variables which are colored and labeled by 5 task factors and 12 survey factors identified by the published exploratory factor analysis. Edges are defined as partial correlations between behavioral performance or self-report measures from 560 participants across all pairs of dependent variables. Edges are sparse due to moderate regularization ($\alpha=0.15$) using Graphical Lasso and a partial correlation threshold of $r\geq0.05$. Nodes are clustered based on a stochastic block model~\cite{peixoto_graph-tool_2014}. Word clouds depict the node identities per cluster. See Appendix \ref{fig2_method} and Ref.~\cite{eisenberg2019uncovering} for details.}
\label{fig2}
\end{figure}

Varied ideas have arisen across fields about the nature of these component processes \textbf{(Figures \ref{fig2} and \ref{fig3})}. Many psychologists view control as emerging from inhibition of automatic or goal-irrelevant responses, updating and monitoring working memory, and shifting attention between tasks~\cite{miyake2012nature, diamond2013executive, bunge2024should}. Cognitive psychologists view control processes as a function of learning to make reward-based decisions by applying selective attention and working memory to monitor and gate ongoing action when there is uncertainty about what to do until sustained or momentary goals are completed~\cite{miller1960plans, rougier2005prefrontal, daw2005uncertainty, koechlin2007information, yee20, braver2012variable, botvinick2014computational, botvinick2015motivation, yee2018computational}. Personality and social psychologists view control as the ability to regulate driving and restraining forces, supported by certain traits and feelings of autonomy, competence, and relatedness, in service of a rich hierarchy of personal goals of differing abstractness, generality, and duration~\cite{deci2000and, carver1982control, duckworth2007grit, duckworth2011meta, kruglanski2012energetics, kruglanski2015architecture}. Expectancy-value theorists in motivation psychology view control as driven by goals that have sufficient value balanced with confidence about the self-efficacy towards expected attainment in service to a system of needs and drives~\cite{atkinson1957motivational, shah1997expectancy, feather2021expectations, lewin2013principles, ryan2002overview, heckhausen2018historical, kruglanski2012energetics}. Neuroeconomists view control processes as a cost-benefit analysis in service of choice that maximizes benefits under costs of alternative opportunities, time, and mental resources~\cite{glimcher2013neuroeconomics, boureau2015deciding, kool2017cognitive, berkman2017self, kool2018mental, lieder2018rational}.

\subsubsection*{Shared computations of dynamic feedback loops}

Despite these differences, all consider the core control process as a dynamic feedback loop of perception, valuation, and action. This feedback loop describes cognitive control as a process balancing the choice to apply control effort with the value and efficacy of that effort, given the perception of the current situation. Choices influence what new sensory inputs, task stimuli, and options one attends to, and the ``perception-action cycle'' continues \cite{fuster1990prefrontal, ochsner2014neural, botvinick2014computational, gross2015emotion, rmus2021role}. Over time, effective control is learning the range of actions that reduce the difference, or increase the congruency, between the expected value of choices and the perceived state of affairs~\cite{miller2017plans, carver2001self, fromer2019goal, bustamante2021learning, mcdougle2022executive}.

Control theory provides systems engineering concepts to understand how dynamic feedback loops operate at multiple scales of time and abstractness for behavioral goals or targets~\cite{carver1982control, powers1973behavior}. Negative feedback reduces the discrepancy between the perceived situation and current goals in order to maintain goal pursuit, while positive feedback amplifies discrepancies to repel further from anti-goals. For instance, thermostats adjust temperature to close the gap with a reference point with a negative feedback loop. Blood clotting involves platelets releasing signals that attract more platelets to the injury site, compounding quickly to create distance from dangerously low levels of blood in a positive feedback loop. 

Many aspects of everyday control behavior involve approaching a target goal or avoiding an anti-goal, in accord with negative and positive feedback loops~\cite{carver2001self}. For example, thinking outside the box involves approaching a creative idea and avoiding familiar and rigid mental sets. Indeed, cognitive computational neuroscience models using supervised learning instantiate positive and negative feedback loops to minimize or maximize multiple cost functions for cognitive control tasks~\cite{miller1960plans, botvinick2001conflict, rougier2005prefrontal,  herd2014neural, kool2017cognitive, ritz2022cognitive}. An effective control system is one that allows for reaching the widest range of routes to goal states efficiently, either in the world or our mind, despite great conflict and given diverse initial contexts and situations. These functional and algorithmic processes guide the investigation of the neural implementation of control.

\subsection*{Neural bases of cognitive control processes}

\begin{adjustwidth}{2em}{0pt}
\textit{``Nothing in excess.''}

\hfill ------ Delphic maxim (circa 6th century BCE)\\
\end{adjustwidth}

A century-old analogy for the neural bases of cognitive control considers how control regions of the brain act as a way station or switch that changes the flow of train traffic across a system of railroads~\cite{james1890principles}. Some of the most influential current conceptualizations of control---the biased competition and guided activation frameworks---apply this analogy~\cite{miller2000prefontral}. The prefrontal cortex can be seen as a central station that implements cognitive control by redirecting trains carrying information about goals to manipulate the flow of command or stop signals at other stations, thereby biasing their activity to produce actions in line with goals~\cite{desimone1995neural, miller2000prefontral}. Control signals coordinate traffic by adding trains, scheduling delays and slowdowns, and selectively admitting or banning only certain types of passengers using selective attention, inhibitory, and learning processes. All these activities help to maximize efficiency and minimize interference, redundancy, and conflict by avoiding traffic congestion as much as possible. There is correlative, causal, computational, and neurodevelopmental support for this framework~\cite{park2009adaptive, somerville2010developmental, parvizi2013will, reuter2014does, cavanagh2014frontal, luna2015integrative, marek2015contribution, cole2017control, widge2019deep}.

\begin{figure}
\begin{center}
 \includegraphics[width=\columnwidth]{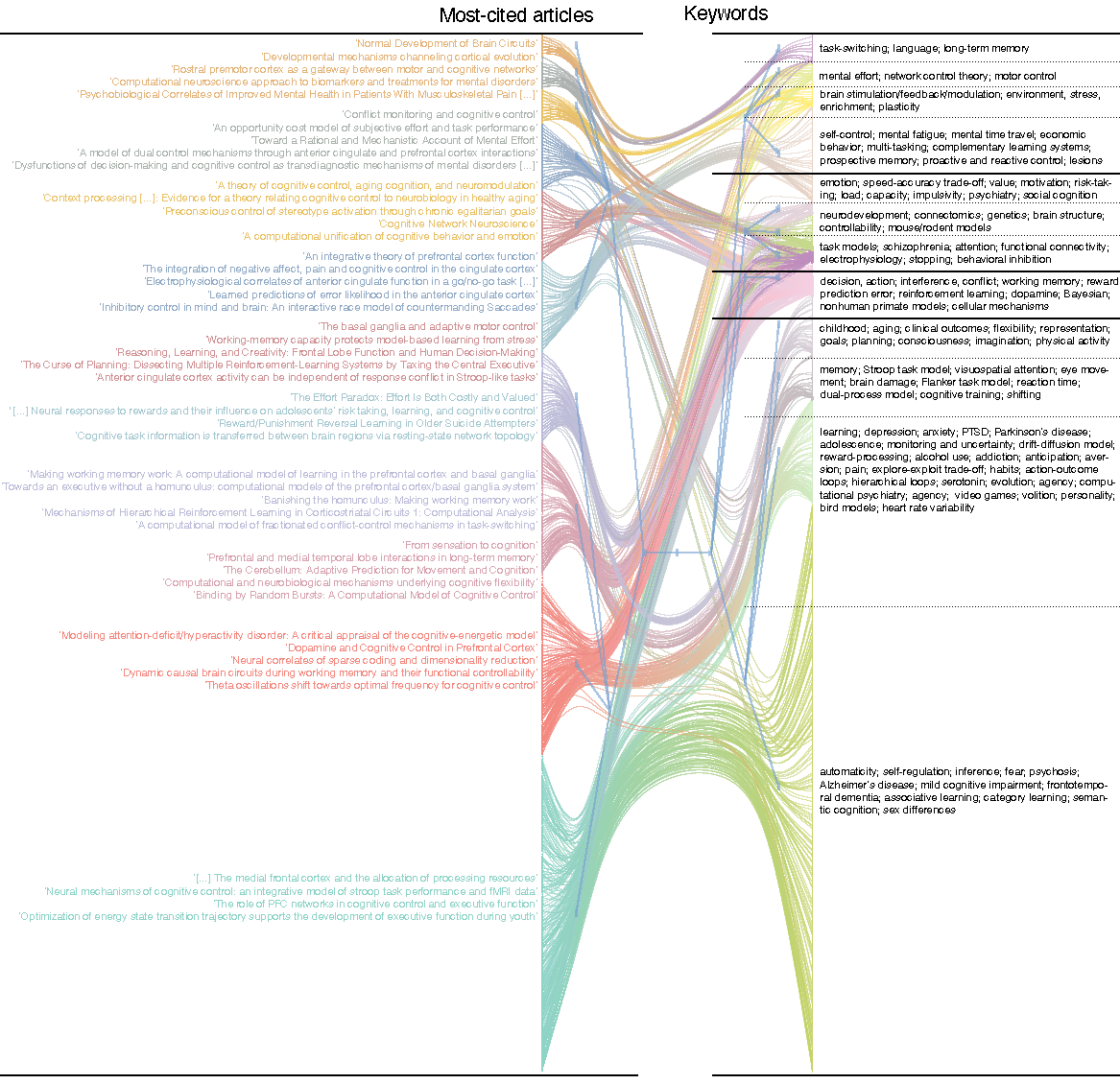}
\end{center}
\caption{\textbf{A plurality of concepts from highly cited articles on cognitive control computations.} The challenge is to develop a cohesive yet precise framework that can speak to a plurality of concepts at multiple levels of analysis. Visualization of a hierarchy of concepts from 308 highly cited papers that included the keywords ``cognitive control'' and ``computational'' from a Web of Science search. Topics determined by a hierarchical stochastic block model of an article-by-keyword co-occurrence network~\cite{gerlach2018network}. For article nodes, we show five or fewer randomly selected articles in the community. For the keyword nodes, we show the most frequent words. Edges indicate co-occurrence. The blue tree network indicates hierarchical community structure, where the highest hierarchical level is demarcated by the solid and bold black lines, the middle level is demarcated by solid black lines, and the most granular level is demarcated by dotted black lines. See Appendix \ref{fig3_method} for details.}
\label{fig3}
\end{figure}

\subsubsection*{Monitoring the conflict and congruency of goals and choices to switch tasks}
Conflict arises from traffic across the brain, or the interference between simultaneously co-activating or spatially overlapping patterns of activity which represent competing or incongruent choices and goals~\cite{monsell2003task, livnat2006optimal, musslick2021rationalizing, fromer2019goal, ritz2024orthogonal, held2024reinforcement}. Traffic makes cognitive control aversive, cost time, feel difficult, and occupy mental resources~\cite{kool2017cognitive, fritz2013conflicts}. The dorsal anterior cingulate and medial prefrontal cortices detect conflict to signal error and to prioritize efforts, sending conflict signals that stimulate resolution among signals with more control~\cite{botvinick2001conflict, narayanan2013common, shenhav2016dorsal}. These efforts are coordinated by goal representations in prefrontal regions, acting as attentional templates to filter or bias downstream neural processing for multiple task rules, goals, opportunity costs, expected values, and levels of effort~\cite{fuster1971neuron, diamond1986comparative, goldman1987s, funahashi1993prefrontal, asaad2000task, miller2000prefontral, badre2008cognitive, badre2009hierarchical, duncan2010multiple, shenhav2013expected, mante2013context, stokes2013dynamic, cole2013rapid, domenech2015executive, freund2021neural, badre2021dimensionality}. For example, dorsolateral prefrontal cortex is thought to bias attention and working memory to promote responses that are appropriate for goals while limiting responses that are inappropriate~\cite{macdonald2000dissociating, miller2000prefontral, friedman2022role}. Ventromedial prefrontal cortex is thought to predict outcomes relative to goals~\cite{duncan1996intelligence, alexander2011medial, miller2000prefontral, shenhav2013expected, friedman2017unity}, regulating arousal and affect \cite{ochsner2005cognitive} as well as adapting to new information that is incongruent with previously learned information \cite{schiller2008fear, moneta2023task}. Ventrolateral prefrontal cortex is associated with the selection of goal congruent response options among competing alternatives~\cite{badre2005dissociable, badre2007left}. These specialized prefrontal regions communicate with a broader frontoparietal and cingulo-opercular network \cite{dosenbach2007distinct, jerde2012prioritized, cole2017control, badre2020brain}, including the anterior insula/operculum, dorsal anterior cingulate cortex, and thalamus~\cite{cocchi2013dynamic, sadaghiani2015functional, wu2019anterior, molnar2022anterior}. Such highly connected regions have many railroad connections, some which make U-turns back to the origin to allow signals to recur stably, that can help to flexibly enact control to modulate conflict, maintain stable control representations, and prioritize activity that is salient or relevant to goal representations~\cite{rougier2005prefrontal, jerde2012prioritized, botvinick2014computational, cole2017control}. As a form of reactive control, competing signals stimulate control signals to resolve conflict, increase resource efficiency, and improve performance by separating previously overlapping representations~\cite{ritz2024orthogonal, braver2012variable, jamadar2015task}. Conflict can also be preemptively avoided by proactively integrating multiple neural representations of the control task and goal~\cite{badre2021dimensionality, kikumoto2020conjunctive, kikumoto2024transient}.

\subsubsection*{Stop signaling and behavioral inhibition}
To alter control and conflict in accordance with goals, some inputs and outputs can be suppressed using the inhibitory signaling of cortico-striatal loops and neuromodulatory pathways~\cite{de2021bi}. Cortico-striatal loops send stop signals from the striatum to cortical regions via thalamic anatomical projections, such as those maintaining information in working memory about goals and the task at hand~\cite{frank2001interactions, wang2004division, frank2006hold, aron2006cortical, aron2007neural, medalla2009synapses, munakata2011unified, aron2014inhibition}. In concert with the inferior frontal gyrus (which includes ventrolateral prefrontal cortex), this fronto-striatal stop circuit inhibits automatized behavior to select actions in pursuit of a goal~\cite{casey2002clinical, liston2006frontostriatal, goghari2009neural, aron2014inhibition}. As such, striatal involvement provides a bottom-up control signal that complicates predominantly top-down views of cognitive control.

\subsubsection*{Value and salience shapes goal-oriented perception, choice, and learning}
Valuation and salience processes also help alter control and conflict activity in accordance with goal representations, calculating which actions to select or suppress in relation to a goal using parallel cortico-striatal loops and neuromodulatory signaling~\cite{daw2005uncertainty, seeley2007dissociable, cools2008role, cools2011inverted, botvinick2014computational, cools2015cost, rmus2021role, mcdougle2022executive}. The ventromedial prefrontal cortex and striatum interact with the dorsal anterior cingulate cortex to support the valuation of the potential courses of action that integrate affective and salience inputs through connections with the anterior insula~\cite{seeley2007dissociable, menon2010saliency, shenhav2013expected, ochsner2014neural, wu2019anterior, molnar2022anterior}. At the neuronal level, incoming and outgoing connections are prone to change such that useful transit lines are more heavily traveled and reinforced by positive feedback or are quieted and closed by negative feedback from other stations, a learning-dependent plasticity that helps save, monitor, and improve progress~\cite{tolman1948cognitive, cohen1990control, o2006making, herd2014neural, o2016leabra}.

\begin{figure}
\begin{center}
 \includegraphics[width=\columnwidth]{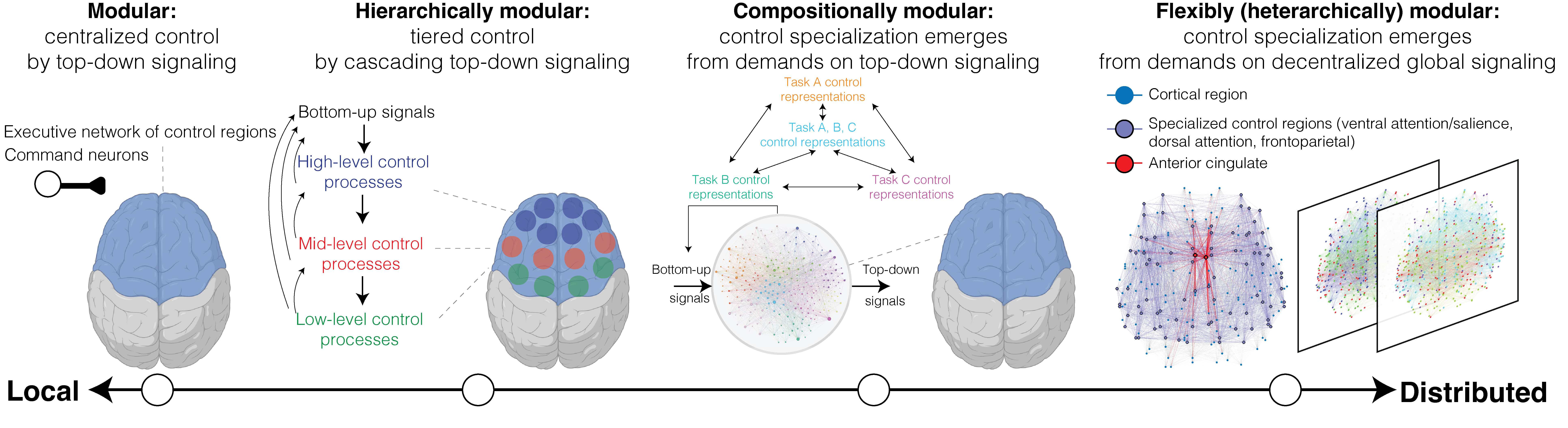}
\end{center}
\caption{\textbf{Local and distributed control signaling}. Hypothesized control processes place differing importance on centralized versus decentralized signaling~\cite{eisenreich2017control}. \textit{Modular control} is largely encapsulated and centralized within a few regions with distinct biological composition to output top-down signals. \textit{Hierarchically modular control} partitions processes along a stratified ladder, maintaining separation of function as top-down signals cascade down the rungs and feed back signals to monitor progress. \textit{Compositionally modular control} emerges from interactively integrable representations learned to meet the demands of control behavior to output top-down and feedback signals. \textit{Flexibly modular control} emerges from representations meeting the demands of control behavior engaged by global top-down, bottom-up, lateral, and recurrent signaling. In contrast to a hierarchy, a heterarchy involves varied modules that consolidate and dissipate over time according to demands on global signaling. Control signals evolve simultaneously and in parallel across neural systems. Which signals lead or follow is relative to the perspective of the scientist (analogous to opening a chain loop to define which link leads and which follows; the break sets which element appears first). These hypotheses can be complementary and need not be mutually exclusive---a heterarchy can involve a temporary hierarchy containing compositional representations in hierarchically nested regions with specialized local biological compositions. Brain image is from BioRender.}
\label{fig4}
\end{figure}

\subsubsection*{Brain-wide top-down and bottom-up signaling and neural signatures of cognitive control}
Theories place differing importance on top-down signaling relative to other forms of signaling~\cite{miller2000prefontral, eisenreich2017control} \textbf{(Figure \ref{fig4})}. Signals of goals, conflict, and stopping are types of top-down signals that filter, amplify, trigger, and redirect certain bottom-up inputs via communication and control~\cite{srivastava2020models}. For instance, bottom-up perceptual signals are biased to select or suppress the parts that are most or least relevant to current goals~\cite{desimone1995neural, noudoost2010top, muller2018cortical}. The collective activity of top-down signals may be enacted by alpha, beta, and theta oscillations from deeper cortical layers, traveling from anterior to posterior regions of the brain~\cite{aron2016frontosubthalamic, cavanagh2014frontal}. These top-down signals may modulate bottom-up gamma bursts from superficial cortical layers to bias perceptual signals as they propagate to task functional regions~\cite{lundqvist2023working, miller2024cognition}. Striatal, insular, and thalamic loops further suggest that motivation, social cognition, reward, and emotions are core signals that dynamically interact with top-down signals, complicating notions of a predominantly top-down hierarchy of control~\cite{deci2000and, pessoa2009emotion, pessoa2023entangled, stalnaker2024neuroscience, leung2024ventral, hofmann2024going}. Recent approaches focusing on neural signatures remain more agnostic to such theories and models of neural systems, and instead classify and decode the patterns of activity across the brain that indicates when a person is actively engaging in cognitive control processes~\cite{schneck2023temporal, rieck2024neural, herzog2025neural}. 

\subsubsection*{Summary}
Control emerges from top-down selective attention or inhibitory gating processes that bias the inflow and outflow of bottom-up signals to filter, amplify, trigger, or redirect parts of the signal related to goal representations. Continuously monitoring the situation, other regions detect conflict between top-down and bottom-up signals to stimulate more control adjustments according to value, cost, and perceived progress.

\subsection*{Limitations and tensions of current psychological and neuroscientific approaches}

\begin{adjustwidth}{2em}{0pt}
\textit{``[T]here occurs a kind of progression in industrial discipline moving from paternalistic controls to assembly line, machine paced routines and, finally, to bureaucratically imposed discipline. What is involved is a shift from heteronomous paternalist controls to autonomous, internalized discipline, and identification with corporate goals and values.''}

\hfill --- Ref.~\cite{o1986disciplinary})\\
\end{adjustwidth}

Current constructs have several limitations and tensions at the neural and behavioral levels, as well as at the cultural and normative levels.

\subsubsection*{Neural and behavioral tensions}

First, it remains unclear exactly how the implicated neural activity represents goals, rules, conflict, costs, benefits, options, actions, and feedback, let alone how different perceptual, affective, motivational, motor, and control representations are integrated with them. In computational models that have incorporated neurophysiological processes and aim to model a cognitive control region's function (e.g. prefrontal cortex), the propagation and transformation of activity allows it to generate cognitive control behavior, dependent on that activity originating from a source that already encodes these kinds of relevant information~\cite{cohen1990control, cooper2000contention, botvinick2001conflict, botvinick2004doing, cooper2006hierarchical, herd2014neural, bustamante2021learning, prystawski2022resource}. Ongoing work aims to explain how cognitive control representations are neurally encoded~\cite{frank2006hold, miller2003recurrent, hazy2006banishing, miller2024cognition, kikumoto2024transient, ritz2024orthogonal, badre2021dimensionality, kikumoto2020conjunctive, van2020goal, yang2019task}. 

Second, what counts as ``good'' cognitive control is ambiguous at the neural and behavioral levels because it depends on context. Different instances of control behavior can be shown to depend on different neural systems. For example, a cognitive control task that focuses on personally or socially relevant goals elicits default-mode activity compared to tasks that focus on more contrived goals~\cite{spreng2014goal, shenhav2013expected}. Moreover, different valuation processes can change what counts as success vs. failure for the exact same control behavior~\cite{mcguire2013rational, kidd2013rational, sutlief2025value}. For example, whether a measure like reaction time, variability, high cost, learning rate, momentary hedonism, or sensation-seeking indicates good or bad control is context-dependent~\cite{mcguire2013rational, kidd2013rational, eisenberg2019uncovering}. This ambiguity arises because many models are underconstrained~\cite{mischel1970attention}, allowing many (sometimes contradictory) ways to explain the same behavior. Fortunately models can incorporate a broader scope of relevant conceptual and neuroscientific variables~\cite{mcguire2013rational, kidd2013rational, lieder2018rational, heffner2021emotion, bustamante2021learning, schurr2024dynamic, silvestrini2023integrative, shenhav2024affective}.

\subsubsection*{Cultural and normative tensions}

Third, a limited set of behavioral instances of control tend to be studied, and those instances seems to cater to societal norms and beliefs about control that are desirable or virtuous. Blending the biology and culture of control blurs the line between self-discipline by internally generated goals and discipline imposed by societal norms~\cite{o1986disciplinary, ryan1995autonomy, deci2013intrinsic}. When construed as encompassing all goal-directed behavior in general, cognitive control can be used to pursue any goal. Yet out of all the countless possible goals that could be pursued, a relatively small set of norms drive outsized attention to goals that seem selected for fitting a certain set of sociocultural expectations about asceticism, mortification (subduing one's ``earthly nature'' and sinful impulses through self-denial), and transgression that can be traced to philosophical and religious literature over two millennia. Striving for temperance, personal responsibility, and meritorious achievement certainly may lead to healthier, happier, and more productive lives and communities~\cite{robson2020self, moffitt2011gradient, duckworth2005self}. It is perhaps unsurprising that endorsement of such norms is related to normative metrics of success that are partly defined by self-surveillance according to those norms~\cite{eisenberg2019uncovering}. But are those goals not more properly value judgments, spiritual exhortations, or moral presuppositions instead of a fundamental neurobiological and cognitive function, even though much could be said in their favor~\cite{duckworth2005self, blair2007relating, moffitt2011gradient, mischel2011willpower, robson2020self}? One wonders how experimental tasks, extant control representations, and outcome metrics of success in life would change if creativity, compassion, cooperation, or altruism were elevated. Just as one could expand the scope of goals and flourishing, conversely, one can readily construe goals wherein strong control leads to guilt, shame, isolation, oppression, deceit, and violence. Or one could construe of goals that appear frivolous, fanciful, or impossible~\cite{chu2024praise}. Normative assumptions reflect the current zeitgiest, but such assumptions not only threaten generalizability across studies and cultures~\cite{munakata2021executive}, but generalizability across time and to other species. There is an ambivalence to cognitive control, more like perception or movement than a virtuous vehicle to prosperity. 

Fourth, cognitive control is not a universal recipe to take action and reap rewards. General applicability to all goals tempts a perennially enticing belief in an inner fuel---like willpower---that can get us to wherever we set our minds to: ``where there is a will, there is a way''~\cite{inzlicht2015six}. In contrast, there is inconsistent evidence for broad applicability of control leading to desired outcomes~\cite{blair2007relating, moffitt2011gradient, duckworth2011meta, burman2015meanings, randles2017pre, eisenberg2019uncovering, enkavi2019large}, evidence for more specific and local transfer across goals based on the type of control strategy and task~\cite{rougier2005prefrontal, collins2013cognitive, duckworth2016situational, duckworth2018beyond,  bustamante2021learning}, and a lack of relationship between control components and putative domain general constructs of intelligence~\cite{friedman2006not}. Willpower is thought to enable mastery over oneself and, with some luck, imposition of one's will on the environment and others. This is power misidentified as willpower. The temptation to confuse power for willpower may be explained by how it empowers people to take responsibility for their fate, a link that justifies why certain people deserve power and others do not, to appear greater than their fortune. One problem that this confusion creates for the science of control is constraining the space of hypotheses to indulge what some may already want to believe---that we can achieve whatever we set our minds to, that we can bend the world to our will, and that clear ambition and focused effort fan fortune towards these ends. These are beliefs about power. When applied to cognitive control, those beliefs support the idea that individual differences in control and their purported outcomes are not only justified but biological. Such beliefs in willpower are outmoded in cognitive control~\cite{inzlicht2019past} and supplanted by dynamic feedback loops of perception, valuation, and action. We do not have worse cognitive control because we lack the strength or capability of will. Rather, worse cognitive control occurs when there is a large discrepancy between what we perceive a situation demands of us and what we have learned to subjectively evaluate as being worth doing, too costly to change, and as opportunities too regrettable to lose given limited time and attention. Nevertheless, a remarkable causal and explanatory power is still often assumed of cognitive control as a solution for the many outcomes that life seems to demand of us, whether or not they arise from our own values. Confidence in this assumption appears disproportionate relative to the construct's modest internal coherence, its modest predictiveness of outcomes with many possible unconsidered causes, and the negative outcomes of its overuse~\cite{smithers2018systematic, eisenberg2019uncovering, munakata2021executive, bustamante2021learning}. 

Finally, many current conceptualizations of control have an issue known as the ``homunculus'' problem, because they postulate internal representations of goals and processes that themselves manage control in the same way a person might. While a homunculus is no longer as explicit as it once was---embodied, for example by Plato's notion of a charioteer of reason or a more recent notion of a central executive CEO who organizes information in their prefrontal office to pursue goals~\cite{plato1952phaedrus, baddeley2002fractionating}---there is still a passing resemblance to human-like controllers in many models, such as an economist evaluating information with rational cost-benefit analysis to maximize the utility of exerting control. Although these metaphorical sources of control are not explicitly posited by any modern theory~\cite{hazy2006banishing, aron2016frontosubthalamic, fuster1990prefrontal, ochsner2014neural, carver2001self}, they are often implied by or necessary to models' formulations, from will to utils. This has been the case since the classic model of dynamic test-operate-test-exit feedback loops, which resembles control by business managers. In a dynamic process with top-down authority and hierarchical communication channels, executive goals set worker objectives, progress is monitored and evaluated, and rewards or punishments are delivered~\cite{miller1960plans, drucker2012practice, simon2013administrative}. Quantifiable goals and task delegation to specialists maximize the efficient distribution of labor and minimize conflicts that prevent resources from being smoothly allocated, strategies with origins in business management~\cite{taylor1919principles, doran1981smart, simon2013administrative}. These are useful and ubiquitous cognitive and cultural tools deserving of application and study, but treating life as if it is a business risks mistaking management metaphors for the nature of self-management. Using metaphors is not necessarily problematic. Metaphors help generate ideas, and they are not necessarily even wrong. After all, humans behave in a world governed by values, traditions, and economics. The issue is when parts of metaphors at the behavioral and psychological level also map out the neural and physical levels. The problem is not that scientists seriously endorse homunculi, but that large knowledge gaps left unexplained about the neural representations of control concepts can implicitly rely on the intuition of homunculi to hammer out and haunt those very voids.  

\subsubsection*{Summary}

We often impose cultural ideas of what control should do onto our models of how control works. The CEO is one such idea, traditions that emphasize rational and virtuous control is another example, and the simplistic idea that people succeed in life because of willpower, where willpower is the currency of control, is another example. The field has already reckoned with many of these nuances, but they have a stubborn influence. It is difficult to cleave ideas that have endured for over two millennia. A neurobiological process model of cognitive control can offer mathematical explanations that help address or circumvent some of these challenges~\cite{eisenreich2017control, pessoa2023entangled, noble2023tip, ross2025explanation}. 

\subsection*{The network control theory construct of control}

\begin{adjustwidth}{2em}{0pt}
\textit{``It is not necessary, however, to organize one's Plans in terms of frozen and brittle terminal states. Unlike good problems in science or mathematics, successful living is not a ``well-defined problem,'' and attempts to convert it into a well-defined problem by selecting explicit goals and subgoals can be an empty deception. [It] is better to plan toward a kind of continual ``becoming'' than toward a final goal.''}

\hfill ------ Ref. \cite{miller1960plans}
\end{adjustwidth}

Originally formalized in the field of systems engineering, control theory has been applied to regulate the behavior of systems in technology, communication, transportation, and robotics \cite{kirk2004optimal, liu2016control, brunton2022data}. In neuroscience and psychology, different forms of control theory have helped to understand perception, movement, reward-guided decision making, and self-regulation~\cite{marken2013perceptual, todorov2002optimal, shenhav2013expected, shadmehr2016effort, shenoy2021measurement, ritz2022cognitive, carver2001self}. Here we describe how network control theory can be applied to understand the neuroscience of cognitive control, at a high level with no assumptions of prior mathematical background.

\subsubsection*{Basics ingredients of network control theory}

As noted above, current models of control in psychology and neuroscience (e.g. the biased competition and guided activation models) describe control regions as implementing top-down signals that have a biasing effect on other regions, guiding their activity as if providing instructions for how to move trains in a system of railroads~\cite{james1890principles, miller2000prefontral, cole2017control}. By contrast, network control theory formalizes these ideas into a cohesive framework by explicitly specifying the kind, amount, location, and timing of activity across the brain expected to enact such control. On this view, an effective control system allows for efficiently reaching the widest range of routes to goal states, given diverse initial contexts and situations. The extant literature on network control theory in neuroscience primarily operates at the level of how control is implemented to efficiently reach the widest range of routes to brain states. 

\begin{figure}
\begin{center}
 \includegraphics[width=\columnwidth]{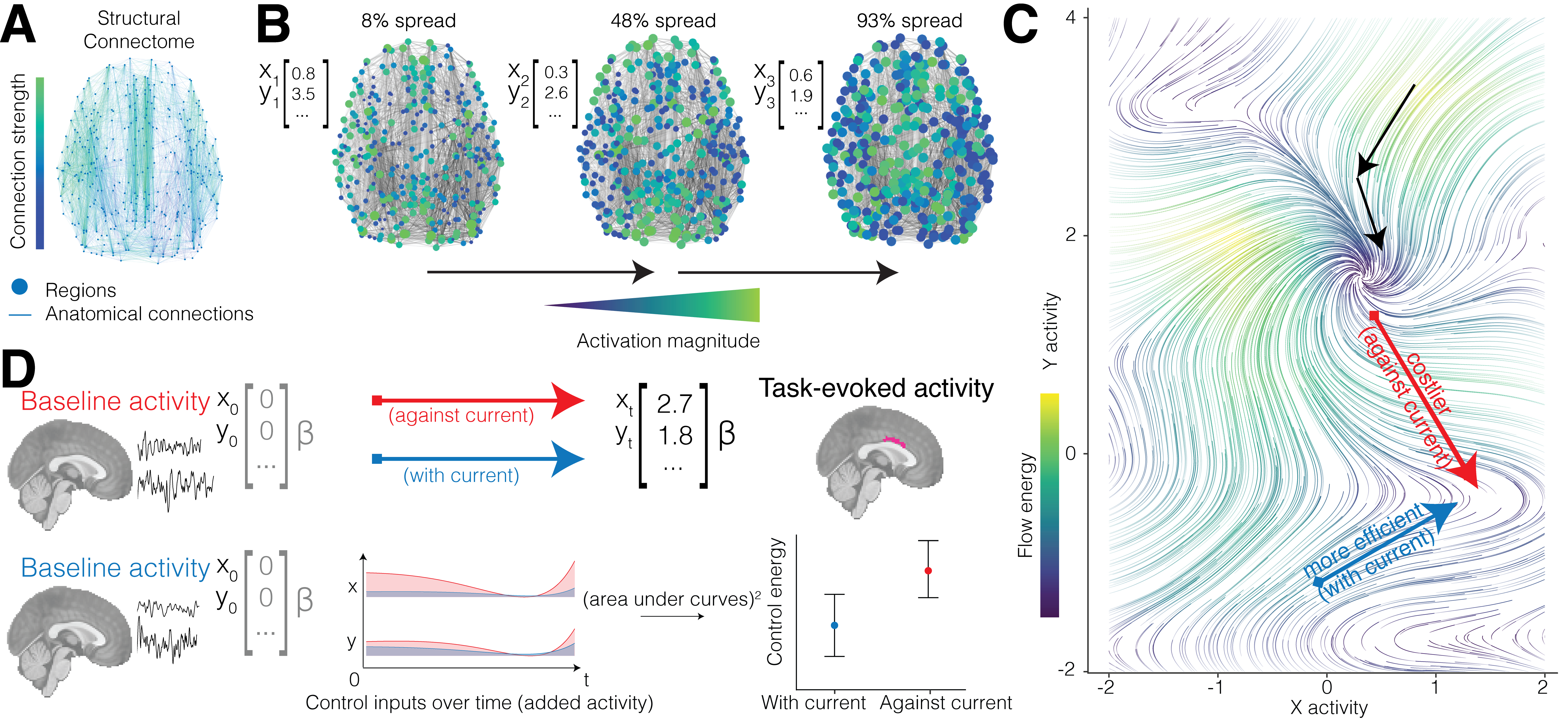}
\end{center}
\caption{\textbf{Guided activation according to network control theory models of spontaneous and controlled activity flow.} \textbf{\emph{(A)}} The anatomical connectivity of the brain, as defined by a structural connectome. \textbf{\emph{(B)}} Neural activity spreads across anatomical connections over time~\cite{avena2018communication, zhou2022efficient, seguin2023brain}. For illustration, consider just two regions (though all are modeled). From an initial state, the change in activity for region $X$ and region $Y$ over two transitions: ($x_1$, $y_1$) to ($x_2$, $y_2$) and ($x_2$, $y_2$) to ($x_3$, $y_3$). The activity values can be predicted linearly, meaning that regional activity is calculated by taking the source activity multiplied by the connection strength and adding it to the target's activity. Linear dynamical systems theory allows us to solve this calculation across all regions and connections simultaneously and understand how global activity patterns emerge from local regional connectivity profiles per time step. \textbf{\textit{(C)}} A conceptual schematic showing how region $X$ and region $Y$'s activity flows over time. The two transitions from Panel \textit{B} are drawn with black arrows. Simulating the predicted activity from many different initial points can provides a comprehensive understanding of the dynamical flow of activity, visualized as streamlines. Streamlines drive activity flow away from zones of higher energy (yellow) to zones of lower energy (blue). The controlled transition between some states (red arrow) must overcome a stronger current of natural dynamics than other states (blue arrow). \textbf{\textit{(D)}} Consider a control problem that starts at two different baselines for the same person. The problem of control is to solve for the minimal time-varying input to add to the natural dynamics to reach the pattern of activity serving as the control target. Here, consider the control target as dorsal anterior cingulate cortex activity evoked during a control behavior that is thought to signal the levels of cognitive conflict and goal congruency. The control energy is the  square of the integrated input added over time such that the energetic cost rises steeply per unit of input. Greater energy is needed for the red trajectory because it must overcome stronger currents of natural activity flow than the blue trajectory. Network control theory provides closed form solutions and tractable numerical simulation tools to find optimal solutions for this problem given any selection of initial states and control targets, and does so simultaneously for any set of regions and connections per time point. Figures reproduced and adapted from \cite{zhou2022efficient} and ~\cite{zhou2023mindful}.}
\label{fig5}
\end{figure}

To illustrate how network control theory works, elements of the railroad analogy ~\cite{james1890principles, miller2000prefontral} remain useful, including the railroads, trains, routes, ongoing traffic, target traffic pattern, and train conductor. In place of railroads tracks, neural signals have the hard constraint of continuously propagating across the anatomical white matter pathways between regions, which can be measured by diffusion imaging \textbf{(Figure \ref{fig5}A)}. Methods to estimate effective connectivity from activity have also been used to model this constraint~\cite{scheid2021time, rouse2021topological}. These ``tracks'' converge at different ``stations,'' or brain regions. 

In place of informational ``trains,'' the \textit{spontaneous dynamics} of each region is determined by propagating threads of activity---signals sent by one region to neighboring regions and still others across their anatomical connections \textbf{(Figure \ref{fig5}B)}. The magnitude of activity is amplified or constrained according to the strength and volume of the connections it travels across. When multiple threads of activity intersect at the same region, their activity is summed, further propagated to connected neighbors, and decays over time.

Due to the hierarchical hub-and-spoke connectivity of the brain~\cite{bassett2010efficient}, what started as hundreds of threads of activity splits off into hundreds of thousands of simultaneous threads as each thread splits off at more and more regional intersections. Due to the small-world modularity of brain structure~\cite{bassett2006small}, after crossing just four or five intersections, threads provide multi-step functional pathways between all regions, although longer functional pathways are far less reliable and more costly to traverse than shorter pathways~\cite{zhou2022efficient, seguin2023brain}. The mathematics used to calculate the predicted future behavior of these dynamics precisely formalizes how much, where, and when neural signals have ``downstream influence'' \textbf{(Figure \ref{fig5}C)}.

The \textit{control energy} is the amount of additional activity that is needed to account for the difference between the future activity predicted in spontaneous dynamics and the amount of future activity that serves as a \textit{control target}. When network control theory is applied to behavioral control tasks, control targets are patterns of activity elicited by a goal or cognitive control task, analogous to prior cognitive and computational neuroscience approaches \textbf{(Figure \ref{fig5}D)}. Defining control targets as future patterns of task-evoked activity can map network control to previously studied activity implicated in the component processes of a cognitive control task (in a standard fMRI analysis, this is a general linear model's beta coefficient). There are also other sensible choices for defining control targets, including measurement units from machine learning classification or decoding methods that index neural signatures of cognitive control, as well as the evolution of neural activity measured with electrophysiology, spike recording, and calcium imaging~\cite{shenoy2021measurement}. The definition of control target is flexible. Modelers can build in but are not bounded by what is considered as a ``correct'' behavior in a well-defined task, providing a useful theoretical foundation for yet unconsidered control targets. 

The amount of energy needed to attain a target pattern of activity can be likened to a calculation of the energy needed to redirect a sailboat towards a target direction by reorienting its sail against the naturally changing currents of wind and water \textbf{(Figure \ref{fig5}C-D)}. The ``energy'' in control energy refers to the cost of additional activity ramping up exponentially. This makes modeled activity evolve over time within relatively stable bounds and prohibits exceptionally large deviations which are not empirically observed in biology due to biophysical constraints~\cite{sterling2015principles}. Despite the different name, energy has the same behavior as the propagating threads of activity described above. Different regions have different available solutions to the control problem of how much, where, and when energy input would usefully add to spontaneous activity to control the evolution of activity to control targets with the allocation of as little control energy as possible. Control energy---just as neural activity and its underlying neurophysiology---does not simply deplete, as posited by resource models of control in psychology whose empirical support has eroded~\cite{inzlicht2019past}. Instead, control energy characterizes continuous readjustments to brain activity by their neural cost, reactively sensitive to inputs such as changing goals or proactively anticipating the demands of the control task in a dynamical prediction feedback loop (``allostasis'')~\cite{ashby1952design, braver2012variable, sterling2012allostasis, zhou2023predictive, theriault2025s, mehrhof2025interoceptive}. 

\begin{figure}
\begin{center}
 \includegraphics[width=\columnwidth]{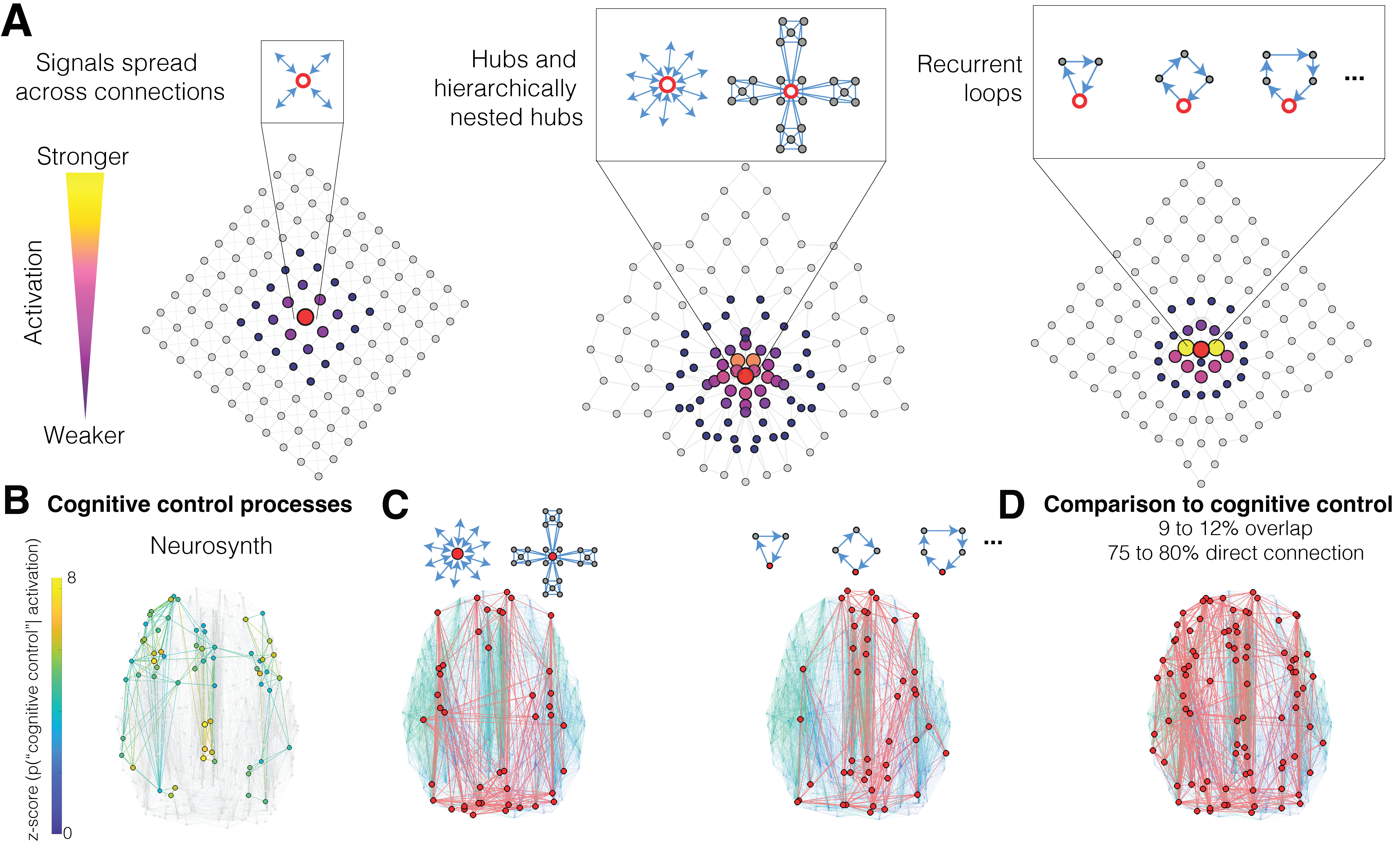}
\end{center}
\caption{\textbf{Nested and recurrent organization of networks shape activity flow.} \textbf{\emph{(A)}} Linear dynamics makes explicit predictions for how connections influence global signal flow for a given region~\cite{avena2018communication, zhou2022efficient, seguin2023brain}. For illustration, let us consider different impulse responses, or how different connectivity profiles for a single source region (red node) influences the spread of activation when the red node receives input. Continuous-time dynamics were simulated using the \textit{nctpy} toolbox~\cite{parkes2023using}. \textit{Left:} When the source is embedded with its neighbors in a grid organization, activity spreads evenly. \textit{Middle:} In contrast, if the source region is a hub---having many more connections than its neighbors---then the activity can spread farther, pulling distant nodes inward to be more reachable by using the hub's many connections. \textit{Right:} If a source region has many more recurrent connections looping back to itself, then activity spreads more than on the grid and in a more concentrated pattern around the source than the hub. Recirculated activity out-competes global spread. These explicit dynamical predictions show how the global activity can be influenced by the differing connectivity profiles of a single source. Network control theory allows us to model multiple simultaneous sources with time-varying input. \textbf{\textit{(B)}} Do regions involved in the cognitive control loops of perception, valuation, and action exhibit recurrent connectivity, suggesting control loops operate at multiple levels? A cognitive control network is visualized, where nodes are highlighted from a Neurosynth meta-analysis of the term ``cognitive control'' with voxels FDR-corrected at $q<0.01$ and regions at $z>3.3$~\cite{yarkoni2011large}. Overlaid on the structural connectome, where edges are highlighted if a direct anatomical connection exists between regions of interest. Gray nodes and edges do not pass statistical thresholds. \textbf{\textit{(C)}} The brain has a mixture of connectivity profiles and multiple regions are potential sources of simultaneous activity flow. Highly connected and recurrent hubs most efficiently distribute control inputs to engage control targets~\cite{betzel2016optimally, gratton2018control, patankar2020path}. \textit{Left:} Hubs with stronger connections than expected by chance comprise a ``rich-club'' of about $12\%$ of all regions. \textit{Middle:} The strongest anatomically recurrent hubs comprise about $12\%$ of all regions. \textbf{(D)} There is an 89\% overlap or direct connection between hubs and the cognitive control nodes, which in combination comprise 27\% of brain regions with highly efficient signaling to the other 73\% of brain regions. Subcortical regions are not included for illustration but see \cite{betzel2016optimally}. See Appendix \ref{fig6_method} for details.}
\label{fig6}
\end{figure}

In place of train traffic causing congestion, interference, and conflict which signal the need for cognitive control, spontaneous activity demands more control energy when it naturally deviates from the control target. As such, cognitive control is not only the consequence of a terminal control target---the activation or deactivation of key brain regions passing some threshold---but is a function of the moment-to-moment processes that guide evolution towards that outcome and continue after that outcome. Network control provides a mathematical model for these continuously evolving functions.

In place of supervision by a metaphorical train conductor or switch operator that follows certain rules and instructions, in network control theory, regions are better network controllers when they mathematically have more possible, or less energy-intensive, routes to control targets. Given the fact that activity continuously flows between regions at all times, good control regions are points in the network where control energy combines with spontaneous activity to effectively distribute the changes in activity levels towards control targets---a dynamic feedback loop constrained by the physics of the brain.

\subsubsection*{Convergence and expansion of predicted network control points with prior regions of interest}

Intriguingly, while all regions could contribute to this continuous and distributed control process, the theory converges on a similar set of regions as prior investigations due to their advantaged placement for adjusting the flow of activity, including the dorsolateral and ventrolateral prefrontal cortex, superior parietal cortex, dorsal anterior cingulate cortex, thalamus, insula, and frontoparietal network broadly~\cite{cui2020optimization, zhou2023mindful, zhou2025neural, miller2000prefontral, gu2015controllability, betzel2016optimally, cole2017control} \textbf{(Figure \ref{fig6})}. Importantly, the theory also naturally produces graded predictions of other key network control points likely to be involved in the dynamic feedback loop of perception, valuation, and action~\cite{betzel2016optimally}, across brain regions and traditionally segregated psychological functions that have been unconsidered, marginalized, or indirectly linked due to variations of task design. These control points include the orbitofrontal cortex~\cite{eisenreich2017control}, caudate and putamen~\cite{lopez2014neural, chiu2017caudate}, hippocampus~\cite{lengyel2007hippocampal, chung2021cognitive, agrawal2022temporal, park2025cognitive}, premotor regions~\cite{fine2022whole}, default-mode regions like the precuneus or posterior cingulate cortex~\cite{spreng2014goal}, visual regions~\cite{powers1973behavior, fuster1990prefrontal}, and amygdala~\cite{pessoa2009emotion, salzman2010emotion, inzlicht2015emotional, pessoa2017cognitive}. 

\subsubsection*{Intuitions about cognitive control from the mathematics of network control}

\begin{figure}
\begin{center}
 \includegraphics[width=\columnwidth]{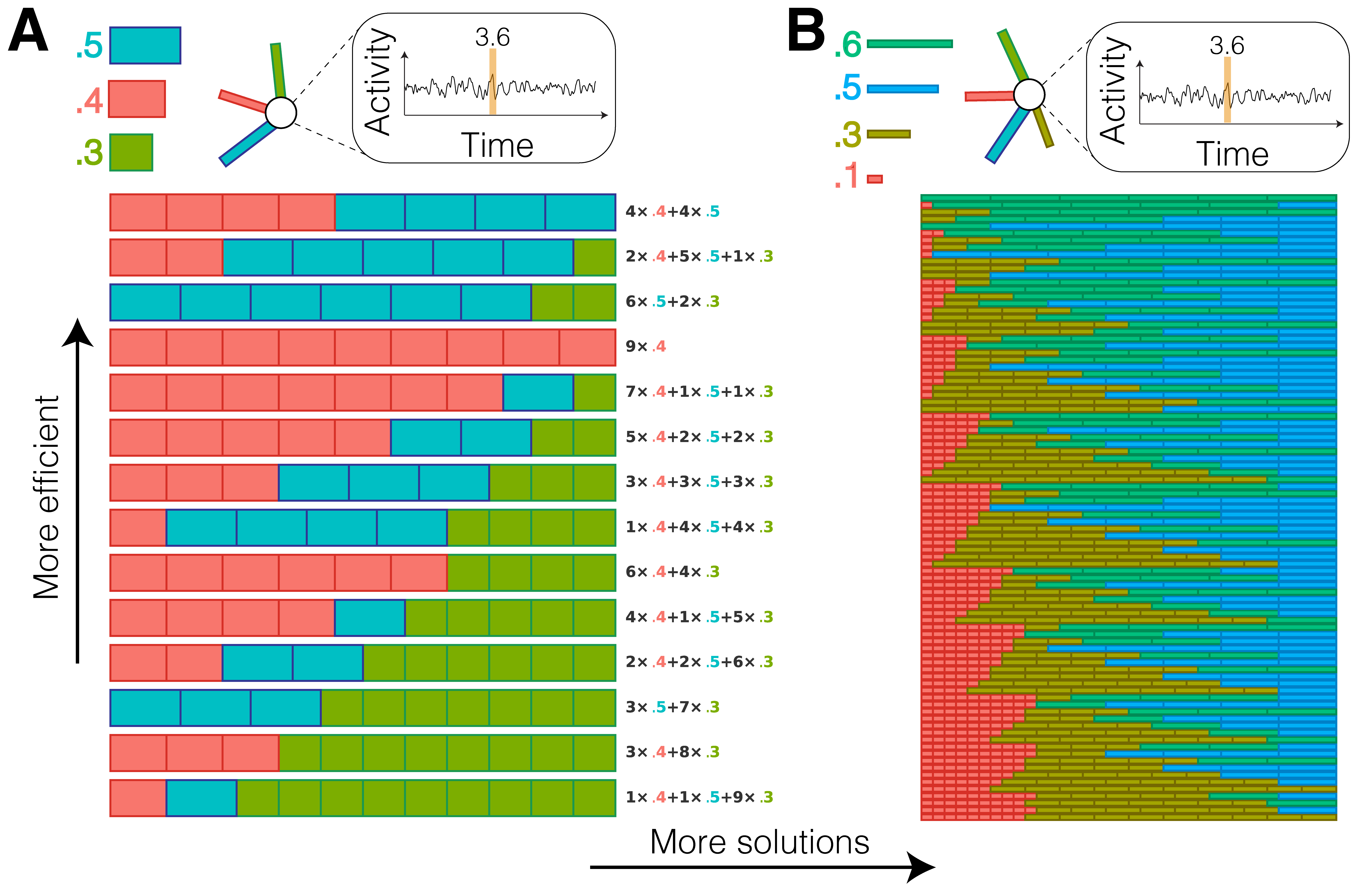}
\end{center}
\caption{\textbf{Making train tracks: developing mathematical intuitions about easier and more efficient control.} \textbf{(A)} Akin to finding all the combinations of railroad tracks in a toy train set that can be used to reach a target destination, the idea underlying the mathematical formalization of control can be illustrated as finding all of the number of ways to make a target pattern (summed activity magnitude of 3.6) with a node (akin to a station) that has three connections of strengths 0.5, 0.4, and 0.3 (where each connection is akin to the specific shape of the railroad track). Each row is 3.6 units of activity wide and specifies one possible solution given the target magnitude and factors. More efficient solutions for reaching the target pattern use fewer connections (akin to the pieces of the track). \textbf{\textit{(B)}} Adding just one more connection vastly increases the repertoire of possible control solutions.}
\label{fig7}
\end{figure}

Having provided a conceptual overview of network control theory, an important next step is formalizing these core ideas with mathematical concepts. Consider the problem of needing to direct activity in a network to control targets, which are the activity patterns associated with a goal or conflict representation. Each target pattern can be represented as a list of activation values $[t_1, t_2, ..., t_n]$ for $n$ brain regions. In the context of the railroad analogy, this process is like figuring out how many ways you can arrange ready-made pieces of railroad tracks of fixed length and curvature to reach different target destinations \textbf{(Figure \ref{fig7})}. This control problem has multiple solutions constrained by the physics of the brain. Each solution is a different way of factoring one of the target activities $t$ as a summed combination of several repeatable factors $x$, $y$, and $z$, which represent the strength of anatomical connections through which activity flows across. Some target destinations are farther than $x+y+z$, necessitating some repeats of each $a$, $b$, and $c$ times, respectively. Taken together, this makes the control problem $t=ax+by+cz$. For example, there are 14 ways of combining the tracks $x=0.5$, $y=0.4$, and $z=0.3$ into the target $t=3.6$ \textbf{(Figure \ref{fig7}A)}. The number of possible solutions to this equation characterizes the probability of control. The quantity $(a+b+c)^2$ characterizes how using a large number of tracks steeply ramps up the energy cost with the squared exponent. Following the previous example, the most efficient solution to factor $t=3.6$ is to combine $x$ four times with $y$ four times for an energetic cost of $(a+b+c)^2=(4+4+0)^2=64$. The most expensive solution combines $x$ once, $y$ once, and $z$ nine times for an energetic cost of $(a+b+c)^2=(1+1+9)^2=121$. Easier and efficient network control has less energetic cost. 

Easy and efficient control is less effortful in the sense of marshalling fewer regions and connections to propagate threads of control energy (activity). For easy and efficient control, it is useful to have $x$, $y$, and $z$ be large because only a few tracks would be needed to make substantial progress towards the target sum. It is also efficient to have a chain of connections that form strong recurrent pathways because an echo of a previous factor can be recycled in a partially diminished form to advance towards the sum without needing to further increase $a$, $b$, or $c$. The advancement persists, saving progress.

Easy and efficient control further benefits from having additional tracks beyond $x$, $y$, and $z$, such as $w$. For example, adding just one more train track $w$ into the mix, so that we have four rather than three pieces available, increases the number of solutions from 14 \textbf{(Figure \ref{fig7}A)} to 89 \textbf{(Figure \ref{fig7}B)}. Having additional tracks is characteristic of hierarchically connected hubs, which are highly connected regions that act as nexuses of activity due to widespread connectivity that forms an information backbone for neural communication~\cite{mesulam1998sensation, van2012high, gratton2016evidence, gratton2018control, badre2020brain, zhou2022efficient, zhou2022compression}. Having more tracks allows for a substantially greater repertoire of control solutions and a larger number of efficient possibilities~\cite{betzel2016optimally, cole2017control}, including stabilizing a current state. Such stability emerges not only from efficiency but from being tethered to more neighbors, aggregating activity changes in many directions and muting rapid fluctuations for lower frequency oscillations of activity that may support information integration for external sensorimotor events, as well as internal decision making, motivation and memory processes~\cite{canolty2010functional}.

In contrast, for difficult and fine-grained control, it is useful to have a diverse range of connection strengths~\cite{kim2018role}. When $x$, $y$, and $z$ are all relatively large, it can create big changes to the sum or a persistent momentum that easily overshoots or strays from the control target pattern of activity $t$. For a greater number of solutions, it is useful to have some small tracks as well, but this comes at a greater cost to $(a+b+c)^2$ from using small tracks many times. Such small additions are more transient, lacking the hefty structural support of hubs to recycle activity. But what is granted is the ability to flexibly turn to new directions with precision, like making a sharp rather than wide turn when driving a compact car versus a truck. For example, a long track travels farther along before a new change in direction is possible, whereas a short track ends quickly allowing for more frequent changes to target destinations. This responsiveness translates to higher frequency oscillations of activity, contributing to faster and more local processing. Oscillations in different frequency bands may support information integration for memory, learning, and attention~\cite{canolty2010functional}.

These intuitions show how network control theory defines control based on the number of ways a region can efficiently utilize connections to dynamically guide patterns of activity to functional control targets. Of those possible network control solutions, several useful metrics assume the usage of \textit{at least} the minimal solution theorized to be sufficient for the control problem at hand and others quantify the number of solutions. This illustration is oversimplified in many ways: for example, solutions must simultaneously account for multiple moving targets with multiple stations, the contribution of factors decay at different speeds, some factors must be used together or in sequence, and solutions are governed by additional spatial and temporal constraints. Nevertheless, the same intuitions apply to more advanced mathematics that allow network control theory to simply but powerfully model how hundreds of regions can simultaneously propagate signals across hundreds of thousands of parallel connectivity pathways over time for any measured or hypothetical control target. Remarkably, as anatomical connections change over development ~\cite{tang2017developmental, lim2025adolescent} and as control targets reconfigure with learning~\cite{betzel2016optimally, braun2021brain, zhou2023mindful}, the brain achieves a unique dynamic reorganization of connectivity and activity for efficient control~\cite{kim2018role}.

There are a number of ways that patterns of activity in a network can be dynamically described by network control theory. This is important, because each pattern has implications for how control is enacted continuously---rather than by a static target pattern of activity---and how it influences the possibility and difficulty of future control such as switching tasks. A control target can be a terminal state that may represent components of a control process, but control theory also determines the persistence and flexibility of the dynamics that lead up to the terminal state and of the dynamics that follow. Persistence is a function of stability granted by low energy, like a ball circling and ultimately resting at the bottom of a bowl. Transience and flexibility is a function of gain granted by high energy, like a ball at the top of a peak ready to roll towards any direction with just the slightest push. The energetic efficiency of dynamics determine the representation's persistence and flexibility, or how resistant or ready it is to changing inputs. Together, network control characterizes dynamical properties of activity, such as energetic costs, competing activity flow, energy peaks, energy valleys, timescales, and optimality of control routes. These properties open the tantalizing possibility of developing dynamical neural models that closely align with the dynamical properties of behavioral and mental control phenomena, such as different levels of subjective effort, conflict, sensitivity to inputs, stable modes, task switching, flexibility~\cite{szymula2020habit, rouse2021topological, cornblath2020temporal, braun2021brain, luppi2023transitions, zhou2023mindful}.

\subsubsection*{Applications and evidence for network control theory}

Network control theory posits mathematical constructs that have been shown to be associated with psychological and neurobiological control processes, task performance, and neurodevelopment in a number of ways. For example, regions implicated in attention and control processes have been associated with high capacity for network control~\cite{rouse2021topological, gu2015controllability, cui2020optimization}. Individual differences in network control are correlated with individual differences in performance on cognitive control tasks probing impulsivity, executive function, and working memory \cite{cornblath2019sex, cornblath2020temporal, braun2021brain, cai2021dynamic}. As would be expected if control is costly, top-down activity from frontal regions associated with goal, self-relevance, and control processes uses more control energy than the bottom-up activity of sensory regions~\cite{parkes2022asymmetric}. The amount of cognitive effort demanded by working memory performance and selective attention is related to individual differences in control predicted by network control theory~\cite{cornblath2020temporal, braun2021brain, zhou2023mindful}, and the network control stability of functional levels of activity for working memory is modulated by dopamine signaling at prefrontal regions~\cite{braun2021brain, cools2011inverted}. As would be expected if excessively costly control is less likely to happen, transitions to neural states predicted to be difficult due to their costly control energy demands tend to have reduced probability of occurrence \cite{weninger2022information, cornblath2020temporal} and have reduced frequency \cite{ceballos2024control}. As would be expected if control that is costly feels subjectively effortful, neural states predicted to be difficult due to their high control energy involve more demanding cognitive tasks \cite{braun2021brain, zhou2023mindful}. Finally, the capacity for network control increases with age in conjunction with known periods of neurodevelopment for executive function~\cite{tang2017developmental, cornblath2020temporal, cui2020optimization, lim2025adolescent}. Comparatively, humans have a more efficient organization of anatomical connectivity for network control than drosophila and mice \cite{kim2018role}

Control energy is a mathematical construct centered on activity change, maintenance, and propagation, and as such, it has been associated with myriad aspects of neural signaling, including dopamine \cite{braun2021brain}, serotonin \cite{singleton2021lsd}, GABA \cite{mahadevan2021alprazolam}, and metabolic \cite{he2021pathological, ceballos2024control} signaling. Network control metrics explain the spatial distribution of the genetic expression of excitatory and inhibitory neurotransmitter receptors that have been proposed to support different forms of cognitive control, including dopaminergic, noradrenergic, cholinergic, and serotonergic receptors, which are required to enact diverse neural dynamics for different cognitive tasks \cite{cools2008role, cools2015cost, shine2019human, cools2022neuromodulation, luppi2023transitions}. Together, network control theory offers a mathematical account of cognitive control as an expression of the fundamental architecture of the brain, coordinating the kind, amount, location, and timing of top-down signals to engage activity that is functional for the control tasks at hand. Just as the perception, valuation, and actions of cognitive control operate through dynamic feedback loops, anatomical connectivity itself forms recurrent loops and hierarchically nested hubs that support the efficient propagation of control signals---control loops all the way down.

\subsection*{Limitations and tensions of network control theory}

Network control theory has several limitations. First, the theory adopts some limitations of psychological models when incorporating the same control targets as the evoked neural representations of control. These include limitations about what part of a control process the control target is supposed to represent as well as the types of goals embodied in those targets. However, network control metrics are the product of many ways in which activity evolves to a final state, not only of the terminal state itself. The theory is not wedded to any singular terminal target but the dynamics leading to any known and yet unknown targets that can be multiple realizations of control processes~\cite{badre2020brain, luppi2023transitions}. 

Second, network control theory is currently more biologically reductive than existing psychological, computational, and neuroscience models~\cite{badre2020brain}, but it need not be in the future. Psychological theory speaks more directly to everyday control problems than the neat mathematical solutions for brain dynamics that network control theory presently offers. A preference for functional and algorithmic explanations over ones about neural implementation coheres with the goals of behavioral, cognitive, and computational scientists. Indeed, control behavior is not just the function of deterministic input-output operations of physical machinery~\cite{wiener2019cybernetics}. Control behavior also cannot be reduced to a fully specified and task-specific set of coarse goals, finer plans, and granular actions according to iterative feedback~\cite{carver1982control, cooper2000contention, botvinick2004doing, cooper2006hierarchical, holroyd2018human}. Cognitive control is biopsychosocial, involving the interactions within organisms just as much as with their environment and others~\cite{dayan2012set, gomez2019life, munakata2021executive, hofmann2024going}. Embracing a plurality of different biopsychosocial goals, which lead to seemingly incommensurable approaches, is where the network approach has great potential~\cite{liu2012control}. To this end, network control has already been applied across levels of analysis to psychological state and psychopathological symptom networks~\cite{henry2022control, stocker2023formalizing}. Future work could, for example, use multilayer networks to model biological, psychological, and social dynamics generated by existing cognitive computational models within the network control theory approach.

Third, network control theory can seem both too simple by modeling neural dynamics as a linear system and too complex by overlapping with the simpler network metrics of hubs. The assumption of linear dynamics improves interpretability at the cost of biophysical accuracy, allowing for formal predictions about signal flow that can be transparently translated to the additive and multiplicative contributions of particular regions and the strength of their connections in the brain. Linearity makes it mathematically clear, for example, that network control is strongly intertwined with some prior network metrics by definition, such as the communicability of signals based on the strength of anatomical connections and their resulting contribution to the reachability of control targets~\cite{chen1984linear, cowan2012nodal, betzel2016optimally, cole2017control, gratton2018control, patankar2020path, becker2020network}. Indeed, it has long been argued that control is a function of communication~\cite{ashby1952design, ashby1956introduction, wiener2019cybernetics}, but not that it is merely communication or only a property of communication pathways~\cite{rubinov2022circular}. Beyond the communicability of control signals, both living organisms and machines regulate behavior towards control targets that change over time through continued interaction and according to diverse contexts. Taken together, control targets and architectures determine the cost of network control, when such costs are excessively large, and precisely what control points contribute to those costs (including hubs but also other key regions often left unconsidered). Network control theory adds conceptual and computational tools to integrate and inform complex spatiotemporal interactions of measurement, stimulation, and perturbation~\cite{shenoy2021measurement}, is predictive of more complex non-linear models of neural dynamics \cite{muldoon2016stimulation, kim2020linear, Nozari2020IsTB}, and links the ease of information flow to the efficiency, direction, and localization of control~\cite{zhou2022efficient, parkes2022asymmetric}. However, linear assumptions can come at the cost of both biophysical and computational richness~\cite{cohen1990control, botvinick2001conflict, o2006making, herd2014neural, buss2014emergent, o2016leabra}. Two promising next steps could be applying network control theory to non-linear computational models in order to provide a complementary explanation for what makes an emerging module of computational units become specialized for control during learning~\cite{yang2019task, zhou2022compression, suarez2024connectome} or improving accuracy by using non-linear methods for determining the expected natural flow of activity prior to applying network control theory~\cite{brunton2016discovering, durstewitz2023reconstructing, brenner2024learning}.

\section{Opportunities for expanding, formalizing, and integrating cognitive control constructs}

\subsection*{How network control theory contributes to the psychological understanding of cognitive control}

Network control theory helps synthesize many key ideas in the neuroscience of cognitive control in a precise and mathematically coherent account. For example, it synthesizes the distinction and balance between automatic and controlled activity in service of reaching task-specific activity patterns~\cite{cohen1990control}; the kind, magnitude, and location of guided activation needed for concurrent and parallel control signals to bias top-down and bottom-up systems towards functional target states~\cite{desimone1995neural, miller2000prefontral, badre2008cognitive}; the importance of multiple hierarchically structured control hubs~\cite{cole2017control, gratton2018control, badre2018frontal}; the importance of recurrent connectivity in maintaining stable control representations~\cite{rougier2005prefrontal, botvinick2014computational}; the neurodevelopment of control~\cite{luna2015integrative, marek2015contribution, crone2017neural}; the physical costs of control~\cite{shenhav2013expected, cools2015cost, lieder2018rational, westbrook2020dopamine}; and a general framework for feedback control across levels of analysis~\cite{carver1982control, wiener2019cybernetics, powers1973behavior, miller1960plans}. 

In addition, the extant network control theory literature advances our understanding of cognitive control in several key ways. First, network control theory and conventional brain mapping approaches converge in identifying which regions and networks are most important~\cite{gu2015controllability, cui2020optimization} and how their neurodevelopment may support executive control~\cite{gu2015controllability, tang2017developmental,cornblath2019sex, lim2025adolescent}. Second, several predictions follow from the theory about when and how much control is required given the difficulty of a control task, the probability of activity patterns, resilience to brain damage or disruption, and the effects of perturbations or stimulation~\cite{eisenreich2017control, yan2017network, towlson2018caenorhabditis, cornblath2020temporal, weninger2022information}, providing a useful theoretical testbed for neurostimulation, neurofeedback, and cognitive behavioral strategies. In this way, network-based theories can guide new investigation into regions underappreciated by prior approaches~\cite{cole2017control, yan2017network, towlson2018caenorhabditis}. Third, the approach is not limited to human psychology, neurobiology, and behavior. Less dependent on the values, brain areas, and functions unobserved in other species, network control can also explain control, cognitive flexibility, attention, and learning in non-human primates \cite{szymula2020habit, rouse2021topological}, mice \cite{brynildsen2020gene}, and \emph{C. elegans} \cite{yan2017network, towlson2018caenorhabditis}. Fourth, this approach can provide better empirical predictions of behavioral and clinical variability beyond simpler metrics describing local activity and connectivity \cite{yan2017network, stiso2019white, parkes2021network}. For example,  network control theory can improve understanding of symptoms across the psychosis spectrum~\cite{zoller2021structural, parkes2021network, mahadevan2021alprazolam, dimulescu2021structural}, depression \cite{kenett2018computational, hahn2021network, hahn2023genetic, zhou2025neural}, attention-deficit/hyperactivity disorder \cite{henry2022effect}, substance use \cite{brynildsen2020gene, zhou2023mindful}, bipolar disorder \cite{jeganathan2018fronto, wang2022alterations}, and epilepsy~\cite{scheid2021time, he2021pathological, chari2022drug}. Treatments can be modeled and understood using network control theory, including pharmacological interventions \cite{singleton2021lsd, mahadevan2021alprazolam}, neurostimulation \cite{stiso2019white, beynel2020structural, scheid2021time, fang2022personalizing, wilmskoetter2022language}, and psychological intervention \cite{zhou2023mindful, stocker2023formalizing}. As such, the approach offers unique conceptual and empirical advantages beyond previous network measures of circuit structure and function~\cite{rubinov2022circular, badre2020brain}.

\subsection*{How the psychological understanding of cognitive control contributes to network control theory}

The psychology of cognitive control contributes to network control theory new kinds of behaviors that require control, how those behaviors dynamically change, how they are integrated into daily life, and different kinds of control processes or mental representations. In short, psychological approaches reveal new control targets and  interpretations of network control's mathematical abstractions.

First, psychology can enrich the interpretation of control theory parameters, such as the time horizon, cost function variables, and the locations of energy input. For example, goals are hierarchical and are construed at differing timescales with differing abstractness versus concreteness~\cite{carver1982control, powers1973behavior, braver2014mechanisms}. This motivates the design of control inputs to maintain or retrieve goal representations, over long or short time horizons, located at regions that represent more abstract or concrete information. Incongruency between choices and goals can prolong decision processes from conflict and opportunity cost representations~\cite{bugg2008multiple, fromer2019goal, bustamante2021learning, lee2024revised}, suggesting that behavior is guided by cost functions with functional and representational constraints beyond the spatial and energetic ones already built into control theory models~\cite{betzel2016optimally, dey2021timescale, kim2025inferring}. Second, psychological and cognitive neuroscience research has described multiple sources of feedback that can influence the magnitude and location of control inputs, including feedback from regions processing stop signals, expected value, prediction errors, interoceptive signals, arousal, motivational needs, and emotions. This expanded suite of feedback signals could dampen, amplify, or prolong top-down signals from cognitive control and attention regions \cite{hasselmo2005model, thayer2009heart, barrett2015interoceptive, inzlicht2015six, baiano2021linking, sennesh2022interoception, shenhav2024affective}, and beyond the brain to the body~\cite{thayer2009heart, barrett2015interoceptive, baiano2021linking, sennesh2022interoception}. The regions and neurotransmitters that have been associated with the allocation of control energy overlap with those related to mental fatigue, value, effort, the cost of effort, conflict, and stop signaling~\cite{shine2019human, braun2021brain, zhou2023mindful, luppi2023transitions, zhou2025neural}, suggesting further potential characterizations of control energy~\cite{shenhav2016dorsal, randles2017pre, berkman2017self, kool2018mental, kok2022cognitive}. Third, cognitive neuroscience research shows that ongoing spontaneous activity is not merely noise to be overtaken by control input but is often characteristic of default-mode network processes including prospective and episodic memory, thoughts about the self, and social reasoning~\cite{spreng2010patterns, koban2021self, menon202320}. New constraints could be added to optimized control trajectories to prolong some functional aspects of prior activity or to minimize the cost of returning. Fourth, goal systems theory, social networks, and new proposals of relational reasoning as a component process of control motivate multilayer network control of interconnected brain regions, goals, needs, motivations, and other people to more directly speak to everyday individual and collective control behaviors~\cite{diamond2013executive, kruglanski2015architecture, srivastava2021structural, falk2017brain, hofmann2024going, bunge2024should}.

\subsection*{Future directions} 

\begin{adjustwidth}{2em}{0pt}
\textit{``The examples we used came mostly from domains of achievement and instrumental activity. However, the model isn't just about achievement related affect. Those domains simply provide easy illustrations of the logic. The model is intended to apply to all goal-directed behavior. This includes attempts to attain goals that are amorphous and poorly specified, and for which construing rates of progress isn't easy. It includes goals for which the idea of rates of progress might at first seem odd. Human goals such as developing and maintaining a sound relationship, being a good [parent], dealing honorably and graciously with acquaintances, seeing someone you care about experience happiness and fulfillment, having a full and rich life, even becoming immersed in the flow of fictional lives in a novel or film are amenable to analysis in these terms. These are all qualities of human experience toward which people try to move, goals that evolve or recur across time, as do most goals underlying human action.''}

\hfill ------Ref. \cite{carver2001self}
\end{adjustwidth}

\textbf{Toward a new taxonomy of goals: neural bases and conceptual expansion} How could any one goal be represented, given a vast universe of possible goals? If goals are attentional templates, task-relevant working memory chunks, or memory schemas, they may not have the rapid flexibility and efficiency necessary to integrate and update a wide variety of changing goals---more like moving targets than checkpoints---which draw from long-term memory, needs, and motivations. Network control theory highlights the theoretical importance for control from multiple regions that represent goals and structured memories, including the ventromedial prefrontal cortex, orbitofrontal cortex, and hippocampus~\cite{wilson2014orbitofrontal, schuck2016human, mack2020ventromedial}. These regions may rapidly represent goals as vectors that encode direction, magnitude, timescale, and abstraction~\cite{tolman1951purposive, lewin2013principles, kelemen2016coordinating, bellmund2018navigating, brunec2018multiple, chung2021cognitive, park2025cognitive}, in support of both spatial navigation and more abstract goal-oriented behavior. The vectorial targets may be instances of basic needs, motivations, and drives~\cite{atkinson1957motivational, white1959motivation, bandura1977self, higgins2011beyond} that underlie goal selection~\cite{molinaro2023goal} and shape signals to regions implementing working memory. Applied to cognitive control, these perspectives echo the ``life space'', cognitive map, and vector field metaphor of goals in psychology, with non-spatial elements such as semantic, emotional, social, reward, or motivational features particularly underutilized due to limitations in measurement and modeling tools~\cite{tolman1951purposive, lewin2013principles}. Natural language processing and large language models can now be used to represent complex hierarchies of control problems, needs, and goals as a kind of semantic meaning vector~\cite{pistilli2024civics, bhatia2025computational, tamir2025tracking}. With the tools of network control theory, it is possible to scale beyond the narrow assumption of isolated goals to model the dynamics of more ecologically plausible constellations of interconnected goals and motivations~\cite{kruglanski2009so, kruglanski2018theory, henry2022control, mcgowan2023dense, werner2024motivational}.

\textbf{Social self-regulation and goal-setting through emotion, motivation, mindsets, and narratives.} Goals are dynamic rather than static terminal states, and goal pursuit is interwoven with emotion, motivation, mindsets, and narratives, influencing beliefs about what and how goals can be achieved~\cite{rotter1966generalized, deci2000and, higgins2005value, dweck2019mindsets, dweck2017needs, oyserman2017identity, zacks2020event, molinaro2023goal, werner2024motivational, shenhav2024affective, chen2024causal, chu2024praise}. Beyond an individual's personal goals, some goals have social origins and goal pursuit can be a collective behavior~\cite{falk2017brain, hofmann2024going}. Moving beyond static trait measures, dynamic modeling of individual emotional and motivational states using dense real-world ecological momentary assessment can improve our understanding of cognitive control in context~\cite{mcgowan2023dense, schurr2024dynamic, lee2025precise, tamir2025tracking}. For example, autonomous rather than instructed motivation or goal selection may facilitate goal pursuit and perceived ease by evoking efficient usage of control energy that temporarily stabilizes the focus of the pursuit~\cite{werner2024motivational, molinaro2023goal}. This interconnectivity of domains suggests the need for network representations to track the complex influence among different modalities, combining measurements of psychological states from ecological momentary assessment, performance on multiple tasks, and fMRI within a single framework. Multi-layer network models offer a way to simultaneously represent such neural and psychological influences, capturing how goals and emotions co-regulate behavior across different people, timescales, and contexts \cite{falk2017brain, srivastava2021structural, henry2022control, stocker2023formalizing}.

\textbf{Dynamic, biologically constrained computational models of control.} Recurrent neural networks and other computational models allow researchers to investigate the learning and adaptation of control strategies across tasks, providing insight into how structural connections adapt and give rise to behavioral outputs \cite{cohen1990control, miller2003recurrent, rougier2005prefrontal, herd2014neural, song2016training, musall2019harnessing, yang2019task, jaffe2023modelling, suarez2024connectome}. Network control theory could be leveraged to study how the cost of control shapes the geometry of task representations~\cite{badre2021dimensionality, achterberg2023spatially}, influencing  decision-making under energetic and temporal constraints \cite{shenoy2021measurement}, as well as the conditions that explain why local versus distributed control modules emerge from the communicability of neural networks as they learn~\cite{eisenreich2017control, yang2019task, zhou2022compression, achterberg2023spatially}.

\textbf{Timeliness of interdigitation.} These approaches point toward both theory-rich and ecologically grounded investigations of cognitive control. Now is a momentous period when there is ample neural data at the cellular, genetic, and regional levels, across childhood, adolescence, and in aging, with continuous tracking of behavior, rich methods of disentangling psychological subprocesses using computational models of behavior, and new methods for querying cognitive and emotional states at scale. The observation that this is more data than researchers know what to do with is often used as a call for more theory. Here, we have described one such theory that has accumulated a wide range of evidence in neuroscience, psychology, physics, and bioengineering. We have suggested some possible pathways to bridge the constructs. Towards this end, there are unique opportunities to study cognitive control naturalistically, and to better understand the state dependence thereof in a much more ecologically valid fashion.

\textbf{Conclusion.}  The extant network control theory literature undergirds, synthesizes, prunes, and expands cognitive neuroscience theories of control. For example, network control undergirds and synthesizes the role of distributed regions in guided activation in dynamic feedback loops, prunes cultural and moral assumptions about power or the right kinds of goals, and expands the theoretical scope of controllers and control targets relevant to semantic, emotional, social, learning, memory, and motivational processes. Network control theory is within its first decades of investigation, and in the course of expansion and formalization, it may become clear an entirely different model provides a better fit to psychological and neural data. New measurement modalities may revolutionize how we implement state estimation. The limits of connectomic approaches may not bend to theoretical and numerical approaches to modeling. Practical demands for implementation in medical devices or artificial intelligence may necessitate non-linear rather than linear methods. Psychological models of cognitive control may prove more relevant than neural models to the design of personal interventions, scalable measurement, societal goals, structural changes to public policy, economic systems, and cultural norms. Nonetheless, network control theory offers useful tools to intuitively understand cognitive control as a cohesive dynamic and distributed neural process. A major goal of research is to develop multi-level models and biopsychosocial understanding, because these are more robust, capture more data, speak to diverse goals across disciplines, and offer multiple pathways for intervention. In this review, we have shown how psychological and cognitive neuroscience models of control provide the strongest constraints at the behavioral level---by specifying what kinds of behavior count as instances of control and what kinds of goals might one try to pursue---but the weakest constraints on the brain level. By contrast, network control theory currently provides stronger constraints on the physical level of the brain than it does on the behavioral level. Combining complementary aspects of each could provide the strongest currently available means of theory and modeling across levels of analysis. 

\section*{Acknowledgments}
We thank Dr. Aaron Bornstein for helpful feedback on earlier versions of this manuscript. This research was supported by the Center for Curiosity and was sponsored by the Army Research Office, accomplished under Grant W911NF-18-1-0244. DZ acknowledges support from the George E. Hewitt Foundation for Medical Research. KO was additionally supported by the Center for Brain Mind and Society, ARO grant W911NF-19-1-0411, and grant R01AA023653 from NIAAA. The views and conclusions contained in this document are those of the authors and should not be interpreted as representing the official policies, either expressed or implied, of the Army Research Office or the US government. The US government is authorized to reproduce and distribute reprints for government purposes notwithstanding any copyright notation herein.

\clearpage
\newpage
\newpage
\section{Appendix}

\subsection{Figure 1 Methods}\label{fig1_method}

Figure \ref{fig1} is based on a previously published review in the digital humanities~\cite{burman2015meanings}, which created a concept network for the psychological construct of self-regulation from linked terms in the American Psychological Association Thesaurus of Psychological Index Terms®~\cite{tuleya2007thesaurus}. These terms characterize the semantics of psychological science, defining what the words used by psychologists mean to other psychologists~\cite{vandenbos2007apa, vandenbos2009apa, tuleya2007thesaurus}. 

Following the guidance of the procedure detailed in this previously published review~\cite{burman2015meanings}, we began a search with two terms ``cognitive control'' and ``self-regulation'' in the June 2025 edition of the APA Thesaurus~\cite{APAThesaurusDigital}. These serve as first-level seed terms. We then obtained the hierarchy of concepts linked to the first-level seed terms to generate a second-level of 13 new seed terms. The second-level terms included: cognitive flexibility, executive function (higher order processes), executive functioning measures, set shifting (attentional set shifting, cognitive set shifting), task switching, agency, emotional regulation (affect regulation, expressive suppression), neurofeedback, self-control (willpower), self-Management, self-monitoring (self-observation), self-monitoring (personality), and self-regulated learning. Finally, we obtained concepts linked to the second-level terms to generate a third-level of 38 terms. This procedure produced 67 terms and their associated definitions (range: 1 to 82 words; mean 19 $\pm$ 14 words). The prior work goes one level deeper to fourth-level terms, which is useful because constructs like ``goals'' and ``values'' live at that level, but we choose to stop at terms two links away for concision---the sheer breadth of indirectly related constructs dilutes the original meaning of cognitive control or self-regulation. For the few terms without a definition, we used their list of linked terms as a substitute.

Here, our methods diverge from those of the prior paper in order to use recently developed natural language processing and clustering techniques. To cluster the terms and visualize a network of term similarity, we first sought to normalize the number of words used in each term's definition. To do so, we augment the definitions by a large-language model. Recent work has relied on GPT-4 to write definitions of psychological terms in order to study similarity and ambiguity among constructs~\cite{wulff2025semantic}. Following their example, we prompt for 100- or 150-word definitions from GPT-4o via the API using the prompt: ``You are an expert in psychology and personality theory. Here is a technical dictionary definition of the construct ``\{label\}'': ``\{definition\}''. Expand this into a \{length\}-word version that retains its scientific accuracy while adding conceptual richness and clarity for a research audience. Use formal psychological language.''

Finally, we applied a topic clustering method using a hierarchical stochastic block model~\cite{peixoto_graph-tool_2014}. This method detects community structure in a term-by-definition co-occurrence network~\cite{gerlach2018network}. We construct the bipartite network by defining a biadjacency matrix where rows are the psychological terms and columns are the words used in the definitions. Edges represent the number of times a word appears in a term's definition. 

Using a hierarchical stochastic block model, we can cluster definitions and terms into blocks (topics) without needing to predefine the number of topics or rely on a prior probability distribution like in Latent Dirichlet Allocation. Each topic corresponds to a block of words, forming a distribution over words based on their co-occurrence across documents. Similarly, terms are grouped into blocks that share similar word usage, yielding a distribution over terms for each definition word. These distributions emerge naturally from the inferred block structure inferred. Finally, we visualize the term-by-definition bipartite network and use a unique color for each inferred block. 

\subsection{Figure 2 Methods}\label{fig2_method}

Figure \ref{fig2} uses the data and results from Figures 2, 3, and 4 in Ref. \cite{eisenberg2019uncovering}. After downloading their publicly available data, we followed their procedure to process individual performance on 129 task and 64 survey measures ($n=560)$. We used a $560\times193$ participant-by-measure matrix to create a $193\times193$ measure-by-measure similarity matrix by calculating the pairwise partial correlation between every pair of measures, controlling for the effect of other measures. 

This can be visualized as a fully connected graph, but structure is difficult to ascertain from such a densely connected network. To identify dependencies between the measures, graphical lasso (scikit-learn) applies L1 regularization ($\alpha=0.15$) to make many entries zero and reduce noisy correlations, making the network more interpretable~\cite{friedman2008sparse, pedregosa2011scikit}. This returns a $193\times193$ measure-by-measure similarity matrix, which we threshold to remove very small correlations $|r|<0.05$. We obtain an adjacency matrix similar to the one visualized in Figure 2 in Ref. \cite{eisenberg2019uncovering}. Each node is a measure's dependent variable and each edge is the partial correlation between measurement variables. 

Here, our methods diverge from those in the prior work~\cite{eisenberg2019uncovering}. To find and visualize community structure in this sparse graph, we perform clustering by using a stochastic block model implemented in minimize\_nested\_blockmodel\_dl by the graph-tool library~\cite{peixoto_graph-tool_2014}. This community detection approach relies on minimizing a probabilistic description length objective to reveal nested groupings in the network. In contrast, the force-directed layout used in the original figure relies on spatial and visual heuristics, rather than statistical or generative structure. The community detection procedure reveals that the $193\times193$ matrix is best modeled by 5 communities, naturally separating survey measurements into two interconnected communities and task measurements into 3 interconnected communities. As in the prior work, these regularization techniques reveal that the task measure communities are largely unrelated to the survey measure communities. However, this may be a result of the application of moderate regularization or the extensive abbreviation of the task durations for the sake of practical experimental implementation, larger sample size, and coverage of measurements.

Finally, we visualize the $193\times193$ matrix as a network and color its nodes according to how the variable that the node represents loads onto independent latent factors obtained from the exploratory factor analysis reported in Ref. \cite{eisenberg2019uncovering}. The 129 task variables loaded onto 5 factors in a $129\times5$ loading matrix with the following factors: speeded information processing, strategic information processing,	perception/response, caution, and discounting. The 64 survey measures loaded onto 12 factors in a $64\times12$ loading matrix with the following factors: sensation seeking, emotional control, mindfulness, impulsivity, reward sensitivity, goal-directedness, risk perception, eating control, ethical risk-taking, social risk-taking, financial risk-taking, and agreeableness. The task and survey factor matrices contain loadings $\lambda$, which characterize the strength and direction of the relationship between the variable and an underlying latent factor. 

For each node of the network, we determined the factor with the maximum magnitude loading $|\lambda|$ and color nodes with a unique tone based on their factor names. For each community, we depict the names of the variables corresponding to the community's nodes using word clouds. The size of the words is proportional to its frequency within its community (phrases and names are treated as a single combined word, and then separated after counting their frequency for visualization).

\subsection{Figure 3 Methods}\label{fig3_method}

To create Figure \ref{fig3}, we obtained biliometric data by performing a literature search on Web of Science\textsuperscript{TM} for the keywords, ``cognitive control computational.'' This search resulted in 308 articles. We saved the title, keywords, and abstract in BibTeX format.

Using similar methods as for Figure 1, we again performed topic clustering with a hierarchical stochastic block model. The first step is constructing an article-by-word co-occurrence network~\cite{gerlach2018network}. This is a biadjacency matrix where rows are articles and columns are the words used in the keywords or abstracts. Edges represent the number of times a word appears in an article. 

Several manual and automatic steps were taken for natural language processing. We manually defined and replaced synonymous terms. For example, ``addiction'' replaced the phrases ``substance abuse,'' ``substance dependence,'' ``substance use disorders,'' and ``craving and relapse.'' Known phrases were treated as a single word when calculating frequency. Automatic processing involved stemming and lemmatizing words (e.g., ``studies'' $\rightarrow$ ``study''), as well as removing stop words (e.g., ``the,'' ``is,'' ``and,'' ``in'', and ``of'').

Using a hierarchical stochastic block model, we cluster words and articles into blocks (topics). Each topic corresponds to a block of words, forming a distribution over words based on their co-occurrence across articles. Similarly, terms are grouped into blocks that share similar word usage across articles, giving a distribution over terms for each definition word. This data yielded multiple nested layers of communities, where nodes are grouped into blocks at the lowest level, and these blocks themselves are grouped into higher-level blocks. The block hierarchy is depicted as a tree structure in blue. Finally, we visualize the article-by-word bipartite network, using a unique color for each inferred block, and we only visualize the strongest 1000 article-word co-occurrences.

\subsection{Figure 5 Methods}\label{fig5_method}
To create Figure \ref{fig5}, we reproduced visualizations from Ref. ~\cite{zhou2023mindful} and \cite{zhou2022efficient}.

\subsection{Figure 6 Methods}\label{fig6_method}
To create Figure \ref{fig6}, we query Neurosynth for the term ``cognitive control''~\cite{yarkoni2011large}. We download the association test maps, which contain voxels that quantify the probability of cognitive control reported in studies, given activation at specific voxel coordinates. These voxel values are z-scores from a two-way ANOVA testing for the presence of a non-zero association between term use and voxel activation, controlling for multiple comparison with a false discovery rate (FDR) criterion of $q<.01$. We parcellate the voxels into brain regions according to the Glasser atlas~\cite{glasser2016multi} and visualize the maximum z-score of the voxels within that region so long as it surpasses a threshold of $z > 3.3$ (corresponding to $p<0.001$). Hub and recurrence visualizations are reproduced from Refs. \cite{zhou2022efficient} and \cite{zhou2023predictive}.

\subsection{Citation Diversity Statement}
Recent work in several fields of science has identified a bias in citation practices such that papers from women and other minority scholars are under-cited relative to the number of such papers in the field \cite{mitchell2013gendered,dion2018gendered,caplar2017quantitative, maliniak2013gender, Dworkin2020.01.03.894378, bertolero2021racial, wang2021gendered, chatterjee2021gender, fulvio2021imbalance}. Here we sought to proactively consider choosing references that reflect the diversity of the field in thought, form of contribution, gender, race, ethnicity, and other factors. First, we obtained the predicted gender of the first and last author of each reference by using databases that store the probability of a first name being carried by a woman \cite{Dworkin2020.01.03.894378,zhou_dale_2020_3672110}. By this measure (and excluding self-citations to the first and last authors of our current paper), our references contain 11.07\% woman(first)/woman(last), 12.89\% man/woman, 16.38\% woman/man, and 59.65\% man/man. This method is limited in that a) names, pronouns, and social media profiles used to construct the databases may not, in every case, be indicative of gender identity and b) it cannot account for intersex, non-binary, or transgender people. Second, we obtained predicted racial/ethnic category of the first and last author of each reference by databases that store the probability of a first and last name being carried by an author of color \cite{ambekar2009name, sood2018predicting}. By this measure (and excluding self-citations), our references contain 7.26\% author of color (first)/author of color(last), 10.61\% white author/author of color, 20.06\% author of color/white author, and 62.06\% white author/white author. This method is limited in that a) names and Florida Voter Data to make the predictions may not be indicative of racial/ethnic identity, and b) it cannot account for Indigenous and mixed-race authors, or those who may face differential biases due to the ambiguous racialization or ethnicization of their names. We look forward to future work that could help us to better understand how to support equitable practices in science.

\clearpage
\newpage
\newpage

\bibliographystyle{ieeetr}
\bibliography{./bibfile}

\end{document}